\documentclass[11pt]{article}

\usepackage{jheppub}

\usepackage{cancel}
\usepackage{adjustbox}
\usepackage{mathtools,amssymb,amsthm}
\usepackage[utf8]{inputenc}
\usepackage{enumitem}
\usepackage{xcolor}
\usepackage{graphicx}
\usepackage{multirow}
\usepackage[normalem]{ulem}
\usepackage{longtable}
\allowdisplaybreaks
\usepackage{cancel}
\usepackage{subfigure}
\usepackage{amsmath}
 \usepackage{amsfonts}
 \usepackage{dsfont}
\usepackage{adjustbox}
\usepackage{subfigure}
\newcommand{\be}{\begin{equation}}
\newcommand{\ee}{\end{equation}}
\newcommand{\bi}{\begin{itemize}}
\def \bea {\begin{eqnarray}}
\def \eea {\end{eqnarray}}

\def\ba#1\ea{\begin{align}#1\end{align}}
\def\bad#1\ead{\begin{aligned}#1\end{aligned}}
\def\bg#1\eg{\begin{gather}#1\end{gather}}
\def\bm#1\em{\begin{multline}#1\end{multline}}
\def\bmd#1\emd{\begin{multlined}#1\end{multlined}}

\newcommand{\ignore}[1]{}

\definecolor{darkgreen}{RGB}{50,150,0}

\def \bal#1\eal  {\begin{align} #1 \end{align}}
\def \bga#1\ega  {\begin{gather} #1 \end{gather}}
\def\({\left(}
\def\){\right)}
\def\[{\left[}
\def\]{\right]}
\def\<{\left\langle}
\def\>{\right\rangle}

\newcommand{\eim}{\end{itemize}}
\newcommand{\beq} {\begin{equation}}
\newcommand{\eeq} {\end{equation}}
\newcommand{\bc}{\begin{center}}
\newcommand{\ec}{\end{center}}

\begin{document}

\title{Comments on firewalls in JT gravity with matter}
\author[a]{Chuanxin Cui,}
\author[a]{and Moshe Rozali}
\affiliation[a]{Department of Physics and Astronomy, University of British Columbia, Vancouver, V6T 1Z1, Canada}
\emailAdd{ccxhans@student.ubc.ca, rozali@phas.ubc.ca}



\abstract{We present two discussions of firewalls in JT gravity. First we present an alternative, arguably simpler, derivation of the gray hole conjecture, applying uniformly to all probes of the firewall probability previously discussed. This derivation is based on the wormhole shortening picture using the handle-disk geometry. However we modifies Saad's story utilizing a "Wilsonian" effective gravitational description, adapted to the time scale probed, in which high frequency modes are integrated out generating the gravitational bulk geometries (dual to the genus expansion in the matrix integral side) whereas low frequency modes are more precisely resolved by being represented as eigenvalue D-branes where JT universes can end. This treatment results in an effective "twist factor cutoff" prescription which simplifies the discussion of long time quantities including the firewall probability. In the second part we discuss effects of matter loops on the firewall probability. While such effects lead to new firewall sources, we argue that these matter loop contributions are sub-dominant at late times.}


\maketitle


\section{Introduction and summary}

The black hole information paradox is one of the sharp questions in understanding quantum gravity \cite{Hawking:1975vcx, Hawking:1976ra}. The discovery of BFSS conjecture \cite{Banks:1996vh} and AdS/CFT  duality \cite{Maldacena:1997re,Gubser:1998bc,Witten:1998qj} suggests that information is not lost, as the dual quantum mechanical systems are inherently unitary. However, AMPS has argued that  \cite{Almheiri:2012rt, Almheiri:2013hfa}, the entanglement between old Hawking particles and the vacuum degrees of freedom near the horizon would violate monogamy of entanglement, and thus unitarity and monogamy of entanglement leads to a dangerous firewall behind horizon. In recent years,  significant progress have been made in addressing the firewall paradox \cite{Bousso:2012as, Nomura:2012sw, Maldacena:2013xja, VanRaamsdonk:2013sza,Almheiri:2014lwa, Dong:2016eik, Almheiri:2019hni, Penington:2019kki, Almheiri:2019qdq, Almheiri:2019psf, Penington:2019npb}. Especially, the proposal of "ER=EPR" \cite{Maldacena:2013xja, Susskind:2014yaa} argues that spacetime wormholes formed by entanglement could resolve the apparent violation of entanglement monogamy.

Furthermore, Maldacena suggested a version of the information paradox in the eternal black hole in AdS -- the decay in the two-point function  at very late times is in tension with the expected discrete spectrum of black hole microstates \cite{Maldacena:2001kr}. Recent progress also revealed the deep connections between black hole and discrete chaotic quantum systems \cite{Shenker:2013pqa, Cotler:2016fpe,  Saad:2019lba, Penington:2019kki}. In particular, the two-point function in chaotic quantum systems  oscillates heavily at late time and the average over these fluctuations exhibits "ramp-plateau" structure. On the gravity side, the ramp is understood as a signature of spacetime wormholes \cite{Saad:2018bqo, Saad:2019pqd, Blommaert:2019hjr, Yan:2022nod}.

Recently, Stanford and Yang \cite{Stanford:2022fdt} examined whether these late time spacetime wormholes could give rise to firewalls and connected this phenomenon to a conjecture made by Susskind \cite{Susskind:2015toa}. The "gray hole" conjecture postulates that the typical late time state has equal probability to be in a black hole or white hole state. The initial state prepared by a Euclidean path integral is far from a complexity equilibrium but under an exponentially long Lorentzian time evolution the state is expected to eventually evolves into a typical state and reach a complexity equilibrium. The ensemble of late-time typical states should be time-reversal symmetric, which means that the evolved state should have equal probability to be a black hole state or a white hole state.  This gray hole conjecture is related to firewalls in the sense that for an in-falling observer,  black holes are considered safe due to their expanding interior, which dilutes perturbations, and white holes are dangerous due to their contracting interior, which enhances perturbations. A precise example was given in \cite{Stanford:2022fdt} by applying external perturbations to the thermal field double state.

The main tool used in \cite{Stanford:2022fdt} to investigate the nature of the late time state and its geometry, and thus the gray hole conjecture,  is the "wormhole shortening" mechanism proposed by Saad in \cite{Saad:2019pqd}. The handle-disk geometry in JT gravity allows a spatial slice of two-sided black hole to emit a baby universe and thus effectively return to its earlier state. Stanford and Yang utilized this mechanism to include sufficiently large baby universes, showing that an old black hole has a nonzero probability to tunnel into a white hole, and that the probability increases to 1/2 at Heisenberg time, consistent with the gray hole conjecture. However, the result raises several  questions:
\begin{itemize}
    \item The white hole probability fails to saturate to 1/2 after Heisenberg time.
    \item It is difficult to preserve the normalization condition that the sum of black hole and white hole probabilities is unity.
    \item It is difficult to generalize the result to other circumstances, for example for a single member of the ensemble  dual to  JT gravity, or including dynamical matter.
\end{itemize}

The answer to the first two questions is related to non-perturbative effects. Indeed, in  \cite{Blommaert:2024ftn}, the authors calculated the white hole probability by directly using non-perturbative results from the matrix integral dual to JT gravity. They showed that the probabilities indeed saturate to 1/2 after Heisenberg time and that the probabilities sum up to unity at all times. Similarly, non-perturbative operator constructions also imply a finite tunneling probability \cite{Iliesiu:2024cnh,Miyaji:2024ity}. These approaches use other probes of the late time state, e.g. the two-point function in \cite{Blommaert:2024ftn}.  However, these non-perturbative approaches obscure somewhat the intuitive and geometrical wormhole shorting picture. 

In this paper, we introduce a convenient way to geometrically encode non-perturbative corrections into the wormhole picture. Specifically, starting with a single member of the ensemble, which we refer to as the UV theory, which is  discrete and non-geometrical, we perform the ensemble average in stages, in the spirit of a Wilsonian effective description. The effective description suitable for long time quantities keeps the average energy fixed and organizes modes into "fast" or "slow" by their energy differences (conjugate to Lorentzian time). By integrating out high frequency "fast" modes, in this context all of the eigenvalues of the matrix outside of a fixed microcanonical window, we obtain an effective gravitational description consisting of JT gravity with its usual geometric genus expansions, together with additional FZZT D-branes representing the eigenvalues in the fixed microcanonical window, which are the "slow modes" responsible for the non-perturbative long time effects. This hybrid description retains many of the advantages of the dual geometrical picture, for example the ability to ask inherently geometrical questions, while being adapted to late time quantities.

\begin{figure}[htbp]
  \begin{center}
  \hspace{-2.2cm}
   \includegraphics[width=6cm]{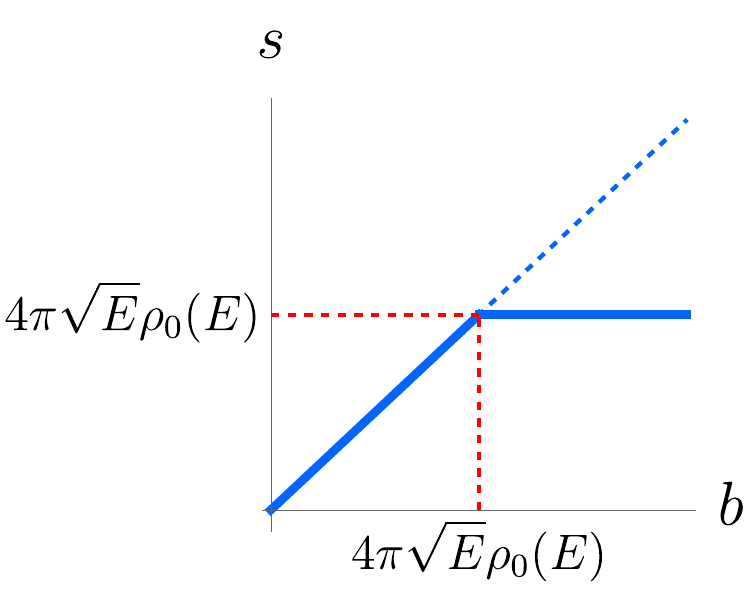}
  \end{center}   
\vspace*{-0.5cm}
\caption{Twist factor cutoff: s is the twist factor and b is the size of baby universe.}
\label{twist1}
\end{figure}

We show then that the effect of singling out the slow modes for a more precise treatment is summarized by a simple modification of the genus expansion in JT gravity. Namely,  the twist factor appearing in the sum over Riemann surfaces is replaced by an "effective twist factor", most significantly having a cutoff arising from treating the FZZT branes exactly. Incorporating this modification into the wormhole shortening picture, this means that small baby universes will continue to form wormholes with the usual twist factor that scales with size of baby universe. But for large baby universes, they will necessarily end on a configuration of D-branes. In that case, the twist factor cutoff accounts for non-perturbative effects and no longer scales with the size of baby universes. We sketch this cutoff in Fig.\ref{twist1}.

Using the twist factor cutoff prescription to encode non-perturbative effects, we repeat the calculation of  wormhole shortening and its effect on the gray hole conjecture. For concreteness we use the two-point function in JT gravity as a probe. This leads to several important differences between our calculation and \cite{Stanford:2022fdt}. First, the bulk moduli of the two-point function  are unconstrained by mapping class group. Secondly, as we discuss below the Euclidean handle-disk geometry contribution to the two-point function is time-symmetric after analytic continuation to Lorentzian times. This means that at late time the black hole can not only emit a baby universe but also absorb one. As a result, the effective time of an old black hole can be reduced or increased. Thirdly, due to twist factor cut off, large baby universes are emitted to or absorbed from
a  D-brane. By carefully including both D-brane and wormhole contributions in a time reversal symmetric manner, we obtain the full non-perturbative result that matches with \cite{Blommaert:2024ftn}:
    \begin{align}
   &\textbf{For }\mathbf{ T< T_H}:\quad \mathcal{P}_{\mathrm{BH}} (T)= 1-\frac{T}{T_H}+\frac{T^2}{2T_H^2}, \quad \mathcal{P}_{\mathrm{WH}} (T)= \frac{T}{T_H}-\frac{T^2}{2T_H^2};
   \label{final-result}
     \nonumber\\
 &\textbf{For }\mathbf{ T \ge T_H}:\quad \mathcal{P}_{\mathrm{BH}} (T)= \frac{1}{2}, \quad \mathcal{P}_{\mathrm{WH}} (T)= \frac{1}{2};
\end{align}
where $\mathcal{P}_{\mathrm{BH}}$ is the probabilities for the late time slice being a black hole state and $\mathcal{P}_{\mathrm{WH}}$ is the probability of being a white hole state, and $\mathrm{T}_\mathrm{H}$ is the Heisenberg time.

Thus, the first two questions mentioned above are answered as:
\begin{itemize}
    \item \textbf{Saturation of 1/2:} After the Heisenberg time, in order to tunnel into white hole all the emitted baby universes necessarily end on D-brane, instead of being traded. This transition leads to saturation of the white hole probability at 1/2.
    \item \textbf{Normalization of probability:} Treating the Euclidean handle-disk geometry in time reversal symmetric way and properly accounting for baby universes ending on D-brane ensures the total probability sums up to unity.
\end{itemize}

The effective geometrical picture we use allows us to discuss some generalizations to pure JT gravity. Suppose for example that instead of integrating over the slow degrees of freedom represented as FZZT branes, we keep those eigenvalues fixed. Similar to the spectral form factor, we see that the twist factor heavily fluctuates (as function of the size of baby universe b). Perhaps these heavy fluctuations could be related to a breakdown of the geometric description or strong back-reaction visible to an infalling observer. However,  we find that since the black (or white) hole probability is an integral over twist factor, the result becomes self-averaging. Indeed, we can go all the way back to the UV picture, namely a a single member of the ensemble, and find that even in that limit the black (white) hole probability is self-averaging. We therefore conclude
\begin{itemize}
    \item \textbf{Individual quantum system:} The gray hole conjecture holds for a single quantum system, provided that the system has a sufficiently large number of degrees of freedom and represents a typical draw from the ensemble that reproduces JT gravity. The black (or white) hole probability is self-averaging.
\end{itemize}

Finally, in the last section we analyze another generalization of pure JT gravity using our effective geometrical description. Namely, we look at the potential source  of a firewall arising from matter loops in the handle-disk geometry, as originally suggested in \cite{Blommaert:2024ftn}. By analyzing the twist mode and the  moduli parameter region corresponding to forming a OTOC contour, we argue that this matter loop effect is sub-leading at late times, suppressed  at least by $\mathcal{O}(1/T)$ before the Heisenberg time and $\mathcal{O}(1/T_H)$ afterwards.

This paper is structured as follows. In Sec.\ref{wormhole shortening}, we review JT gravity and Saad's wormhole shortening picture. In Sec.\ref{FZZT_sec}, we construct the effective description of a microcanonical window involving  JT gravity accompanied by FZZT branes and calculate the cutoff on the effective twist factor. In Sec.\ref{grayhole_sec}, we use the twist factor cutoff to calculate the probabilities of effective time slice using handle-disk geometry and make some comments on the results. In Sec.\ref{matter-loop_sec}, we examine the effect of matter in the wormhole background and argue that the matter loop correction does not lead to a firewall.

\vspace{10pt}
\section{Wormhole shortening in JT gravity}
\label{wormhole shortening}

In this section we briefly review JT gravity, mainly for the purpose of explaining our notations and conventions. For more details we refer interested reader to overviews in e.g.  \cite{Sarosi:2017ykf, Trunin:2020vwy, Mertens:2022irh, Jafferis:2022wez}.   We then summarize Saad's perspective on the "wormhole shortening" effect in the handle-disk geometry \cite{Saad:2019lba} which corresponds to the process of baby universe exchange. In the following section we argue that this picture is modified for sufficiently large baby universes and present a more complete picture.

\vspace{5pt}
\subsection{JT gravity setup}

JT gravity is a 2d dilaton gravity model with a special choice of linear dilaton potential:
\begin{equation}
    I_{\mathrm{JT}}=-\frac{S_0}{2\pi}\left[ \frac{1}{2}\int_{\mathcal{M}}\sqrt{g}R+\int_{\partial\mathcal{M}}\sqrt{h}K \right]-\left[ \frac{1}{2}\int_{\mathcal{M}}\sqrt{g}\phi(R+2)+\int_{\partial \mathcal{M}}\sqrt{h}\phi (K-1) \right],
\end{equation}
where $\mathcal{M}$ is a two-dimensional manifold with Euclidean metric $g$, $h$ is pull-back metric at the boundary $\partial\mathcal{M}$, $S_0$ is the extremal entropy (appearing as a free parameter in JT) and $\phi$ denotes the dilaton field. The first bracket is a topological invariant proportional to the Euler characteristic: $-S_0\mathcal{X}$, and the dynamics comes from the second bracket. 

Below we will be interested in boundaries in JT gravity. There are in total four inequivalent boundary conditions \cite{Goel:2020yxl}. For now we  will mostly be interested in the Dirichlet-Dirichlet (DD) boundary condition, which corresponding to asymptotically $AdS$  boundaries, such that the regulated boundary length is fixed to be $\frac{\beta}{\epsilon}$ and the dilaton value at boundary is $\frac{1}{2\epsilon}$. When discussing D-branes below, We will also consider Dirichlet-Neumann (DN) and Neumann-Dirichlet (ND) boundary conditions studied in \cite{Goel:2020yxl}.

 After integrating out dilaton field, the bulk of JT gravity is frozen to be locally $\mathrm{AdS}_2$ and the only dynamics comes from boundary gravitons, which can be described as a one-dimensional Schwarzian theory with a density of states at disk level \cite{Cotler:2016fpe, Stanford:2017thb, Yang:2018gdb}:
\begin{equation}
     \rho_0 (E) =  \frac{e^{S_0}}{2 \pi^{2}} \sinh{\left( 2 \pi \sqrt{E}\right)}.
    \label{dos_JT}
\end{equation}

The two-sided Hilbert space (dual to  the eternal black hole) can be described in different bases, related by different resolution of the identity operator \cite{Yang:2018gdb, Saad:2019pqd}:
\begin{align}
     \mathds{1} &= \int_{0}^{\infty} dE \rho_0 (E) |E\rangle \langle E| = e^{-S_0}\int_{-\infty}^{\infty} d\ell |\ell \rangle \langle \ell|,
                \label{identity_opt}
\end{align}
where $|E\rangle$ is the energy basis, $|\ell \rangle$ is the length basis with renormalized length $\ell$ of a spatial slice\footnote{The renormalized length $\ell$ is related to the  divergent length $\hat{\ell}$ of a spatial slice by $\ell=\hat{\ell}- 2\log \frac{1}{\epsilon}$, this allows $\ell$ to take negative values.}. 
These bases obey the orthogonality relations:
\begin{align}
   e^{-S_0} \int_{-\infty}^{\infty} d\ell \langle E | \ell \rangle \langle \ell | E^{\prime} \rangle &=\frac{\delta(E-E^{\prime})}{\rho_0 (E)},
   \label{orth_E}
   \\
     \int_{0}^{\infty} dE \rho_0 (E)  \langle \ell | E \rangle \langle E | \ell^{\prime} \rangle &= e^{S_0}\delta(\ell-\ell^{\prime}),
     \label{orth_l}
\end{align}
The Wheeler-Dewitt wavefunction $\langle \ell | E \rangle$ is given as \cite{Harlow:2018tqv,Yang:2018gdb}:
\begin{align}
     \psi_E(\ell) \equiv \langle \ell | E \rangle =2 K_{2i \sqrt{E}}(2 e^{-\ell/2}).
     \label{WD}
\end{align} 
One can then express bulk Hartle-Hawking state $|HH_{\beta/2}\rangle$ with boundary length $\beta/2$ in terms of energy basis:
\begin{equation}
    |HH_{\beta/2}\rangle=\int_0^{\infty} dE \, \rho_0 (E) \, e^{-\frac{\beta}{2}E} |E\rangle.
\end{equation}

By using (\ref{WD}), the Hartle-Hawking state can also be expressed in the length basis. The disk partition function of JT gravity can be interpreted as the inner product of two Hartle-Hawking states \cite{Saad:2019pqd}:
\begin{equation}
     Z_{\mathrm{disk}}(\beta) =\langle HH_{\beta/2}|HH_{\beta/2}\rangle 
         =\frac{e^{S_0}}{2 \sqrt{\pi} \beta^{\frac{3}{2}}} e^{\frac{\pi^2}{\beta}}.
         \label{disk_parti}
\end{equation}
More generally, higher topologies with multiple asymptotic boundaries can be obtained from surfaces with geodesic boundaries by gluing a trumpet geometry, which has an asymptotic boundary of length $\beta$ and  a geodesic boundary of length $b$, for each such boundary \cite{Saad:2019lba}. The moduli space of surfaces of fixed geodesic boundaries is given by Weil-Petersson volume $V_{g,n}(b_1,...,b_n)$, then partition function for genus g surfaces with asymptotic boundaries is given by,
\begin{equation}
    Z_{g,n}(\beta_1,...,\beta_n)=\int_0^{\infty} b_1 db_1...\int_0^{\infty} b_n db_n V_{g,n}(b_1,...,b_n) \psi_{\mathrm{Tr}}(\beta_1,b_1)...\psi_{\mathrm{Tr}}(\beta_n,b_n),
    \label{muti_bdy_pati}
\end{equation}
 where the  trumpet contribution is denoted as $\psi_{\mathrm{Tr}}(\beta,b)$, $n$ is the number of asymptotic boundaries and $g$ is the genus. The measure factors $b_i$ are included due to possible twists in the gluing process and they become very important for the late time behavior of many quantities, as we discuss further below.

Another perspective on the above description is to treat the addition of a  geodesic boundary $b$ as an adding a closed baby universe state, interpreted in a third quantized language. The above trumpet geometry can then be expressed as the following wavefunction in the third quantized Hilbert space \cite{Yang:2018gdb,Saad:2019pqd}:
\begin{align}
    \psi_{\mathrm{Tr}}(\beta,b) &\equiv  \langle HH_{\beta/2} | b,HH_{\beta/2}  \rangle = \int dE  \frac{\cos (b\sqrt{E})}{2\pi \sqrt{E}} e^{-\beta E}.
    \label{asym_baby}
\end{align}

\begin{figure}[htbp]
    \centering
    \subfigure[]{\includegraphics[width=0.26\textwidth]{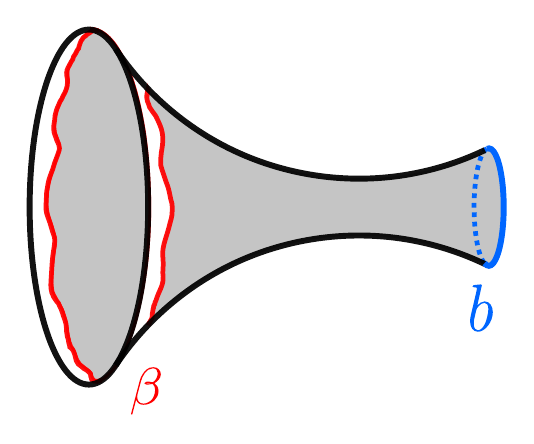}} 
    \subfigure[]{\includegraphics[width=0.26\textwidth]{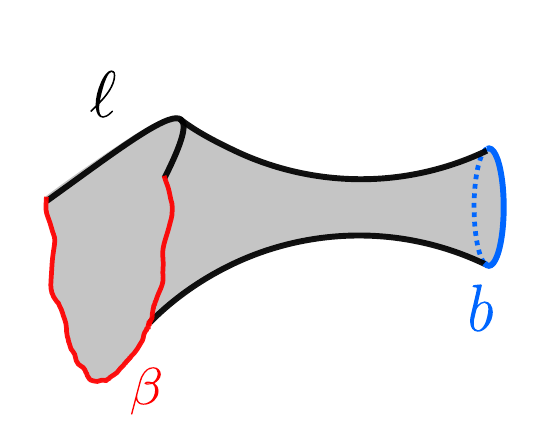}} 
    \subfigure[]{\includegraphics[width=0.26\textwidth]{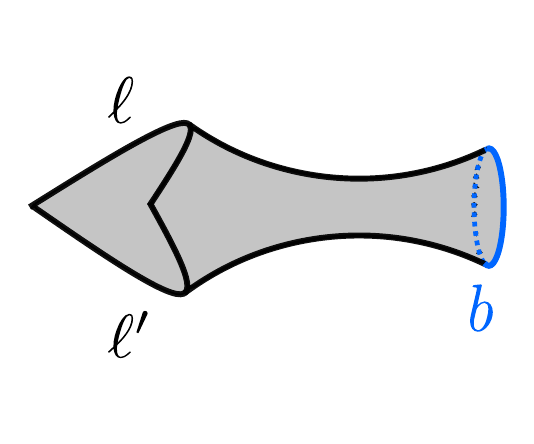}}
    \caption{(a) asymptotic boundary $\beta$ with geodesic boundary $b$: $\psi_{\mathrm{Tr}}(\beta,b)$; (b) mixed asymptotic boundaries $\beta,\ell$ with geodesic boundary $b$: $\psi_{\mathrm{Tr},\beta/2}(\ell,b)$;  (c) asymptotic boundaries $\ell, \ell^{\prime}$ with geodesic boundary b: $\langle \ell^{\prime} | \ell,b \rangle$ }
    \label{trump_WF}
\end{figure}

By inserting the resolutions of the identity in (\ref{orth_E}) and (\ref{orth_l}) into the above expression, one can also obtain the following trumpet wave functions:
\begin{align}
     \psi_{\mathrm{Tr},\beta/2}(\ell,b) &\equiv \langle HH_{\beta/2} | \ell,b \rangle = \int_0^{\infty} dE \frac{\cos(b\sqrt{E})}{2 \pi \sqrt{E}}e^{-\frac{\beta}{2}E} \psi_E(\ell) ,
    \label{asym_l_baby}
    \\
     \langle \ell^{\prime} | \ell,b \rangle &= \int_0^{\infty} dE\frac{\cos(b\sqrt{E})}{2 \pi \sqrt{E}} \psi_E(\ell) \psi_E(\ell^{\prime}) .
    \label{l_baby}
\end{align}
We draw the geometric pictures of above three kinds of trumpet wave functions in Fig.\ref{trump_WF}.

\vspace{5pt}

\subsection{Including matter}

We will also consider JT gravity minimally coupling to a free massive scalar field by adding the following action:
\begin{equation}
    I_m=\frac{1}{2}\int_{\mathcal{M}}\sqrt{g}\left( g^{ab} \partial_a\chi \partial_b \chi +m^2 \chi^2 \right) ,
\end{equation}
where the matter field $\chi$ couples to the metric but not to the dilaton. We take the Dirichlet boundary condition so that the dual boundary operator $\mathcal{O}$ has scaling dimension\footnote{Neumann boundary condition changes the quantization such that the sign in front of square roots in (\ref{dim-mass}) are flipped.}:
\begin{equation}
    \Delta=\frac{1}{2}+\sqrt{\frac{1}{4}+m^2} .
    \label{dim-mass}
\end{equation}

We are interested in the Lorentzian correlator of two-sided boundary operators in the probe limit, meaning that back-reaction effects are ignored. This two-point function at Lorentzian time T is defined as:
\begin{align}
    G_{\Delta}(T) \equiv \frac{1}{Z(\beta)} \langle HH_{\beta/2}|e^{iHT}\mathcal{O}_{L} \mathcal{O}_R e^{-iHT}|HH_{\beta/2}\rangle.
    \label{def_2pt}
\end{align}
where we use in the definition the bulk Hartle-Hawking state creating the two sided black hole at $T=0$ and bulk two-sided Hamiltonian $H_L+H_R$ which propagates both left and right boundary operators in time (the combination $H_L-H_R$ is a symmetry).

At the disk level, the boundary propagator is \cite{Yang:2018gdb}:
\begin{align}
    \langle \mathcal{O}(x_1) \mathcal{O}(x_2)\rangle_{\mathrm{disk}}=e^{-\Delta \ell(x_1,x_2)},
    \label{disk-2pt}
\end{align}
where $\ell(x_1,x_2)$ is the renormalized length of  the bulk geodesic  connecting the boundary locations $x_1$ and $x_2$. Thus the two-point function on the disk geometry can be easily calculated by inserting an identity operator expressed in the length basis, which yields:
\begin{align}
    G_{\Delta}^{\mathrm{disk}}(T)=\frac{e^{-S_0}}{Z(\beta)}\int_{-\infty}^{\infty}d\ell \langle HH_{\beta/2+iT}|\ell\rangle \langle \ell|HH_{\beta/2-iT}\rangle e^{-\Delta\ell}.
\end{align}


We present this formula geometrically as below:
\begin{align}
    G_{\Delta}^{\mathrm{disk}}(T)&= \adjincludegraphics[width=4.5cm,valign=c]{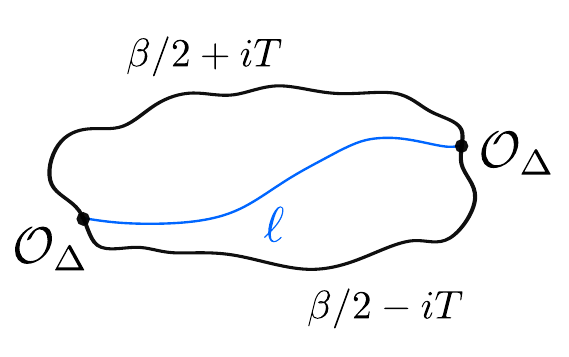},
    \\
    &=\frac{e^{-S_0}}{Z(\beta)} \int_{-\infty}^{\infty} d\ell \adjincludegraphics[width=4cm,valign=c]{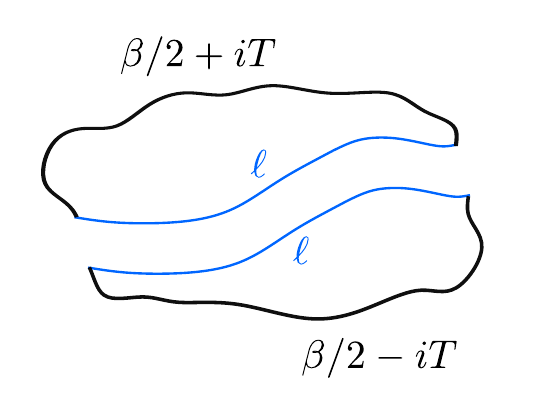} e^{-\Delta \ell}
\end{align}

Two-point function on higher genus geometries requires summing over all inequivalent geodesics in the bulk. We will calculate the genus one case in detail in Sec.\ref{grayhole_sec}.

\vspace{5pt}
\subsection{Wormhole shortening}

In \cite{Saad:2019pqd}, Saad established a nice "wormhole shortening" picture for understanding the ramp structure in the spectral form factor (SFF) and two-point functions at late time (in the next section we further discuss the structure of the plateau). Here we review the picture in the context  of the SFF, for full details we refer to the original paper \cite{Saad:2019pqd}.

The SFF can be reformulated as the "return probability":
\begin{equation}
    |Z(\beta+iT)|^2=|\langle HH_{\beta/2}|e^{-iT H}|HH_{\beta/2}\rangle|^2.
    \label{return_amp}
\end{equation}
This is the probability for the Hartle-Hawking state to return to itself after a Lorentzian evolution of time T. The states in JT gravity can be characterized by the length of the Einstein-Rosen bridge (ERB). Therefore, the magnitude of the SFF quantifies the probability for the length of ERB at late time to go back to its initial value.

\begin{figure}[htbp]
  \begin{center}
   \includegraphics[width=0.4\textwidth]{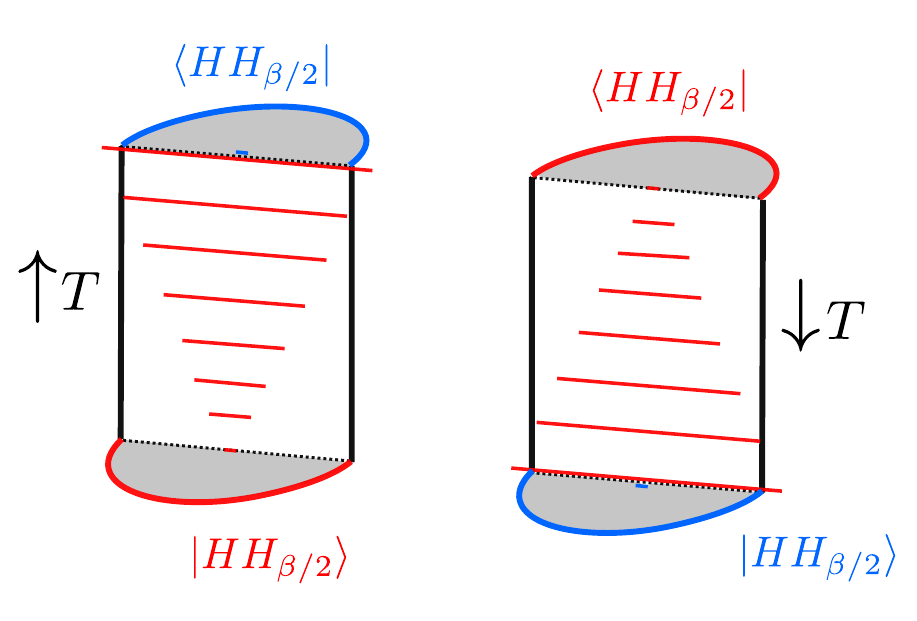}
  \end{center}   
\vspace*{-0.5cm}
\label{disk_SFF}
\caption{Disk contribution to the SFF. Red and blue lines represent the ERB, with their length qualitatively corresponding to ERB length.}
\end{figure}

The leading order contributions to the SFF at early time are two disconnected disks. The size of the ERB of at Lorentzian time $T$ can be obtained by calculating the expectation value of the length operator $\ell$ in the Hartle-Hawking state. The semi-classical result shows that ERB grows linearly at late times \cite{Yang:2018gdb}:
\begin{align}
    \mathcal{V}(t)= \frac{e^{-S_0}}{Z(\beta)}\int_{-\infty}^{\infty}d\ell \langle HH_{\beta/2+iT}|\ell\rangle  \langle \ell | HH_{\beta/2+iT}\rangle \ell \approx \frac{2\pi T}{\beta}, \quad T\gg \beta,
    \label{ERB_length}
\end{align}
where $\beta$ has being analytically continued to $\beta+iT$ to account for the Lorentzian evolution of the Hartle-Hawking state: $|HH_{\beta/2+iT}\rangle = e^{-iTH} |HH_{\beta/2}\rangle$.  Since the length of the ERB is growing linearly with respect to Lorentzian time, the return amplitude decreases, giving the early time "slope". We illustrate the Lorentzian picture for the disk topology in Fig.\ref{disk_SFF}.

At late enough time when the disk contribution decays to be sufficiently small, the next order topology, the handle-disk, dominates the SFF. This topology allows the two systems to exchange a baby universe with arbitrary size, meaning that the length of ERB can be reduced or enlarged by emitting or absorbing a baby universe. It turns out that the length is approximately conserved in this process, $\ell \sim \ell_T \pm b$, where $b$ is the size of baby universe and $\ell_T$, $\ell$ represent the sizes of the ERB before and after the exchange process, and $\pm$ signs stands for the possibility of emitting/absorbing a baby universe. We see then that if the size of an emitted baby universe is approximately $b \sim T$, the baby universe exchange reduces the length of ERB to its original length at $T=0$ and thus provides a non-decaying amplitude. 

Importantly, the exchange process involves gluing of two-baby universes emitted from both copies of the system with an arbitrary twist. This twist measure factor is responsible for the appearance of a linear in $T$ ramp. This mechanism is also responsible for the appearance of a ramp of two-point function (and higher correlation functions) at late times \cite{Saad:2019pqd}. 

\begin{figure}[htbp]
    \centering
    \subfigure[]{\includegraphics[width=0.4\textwidth]{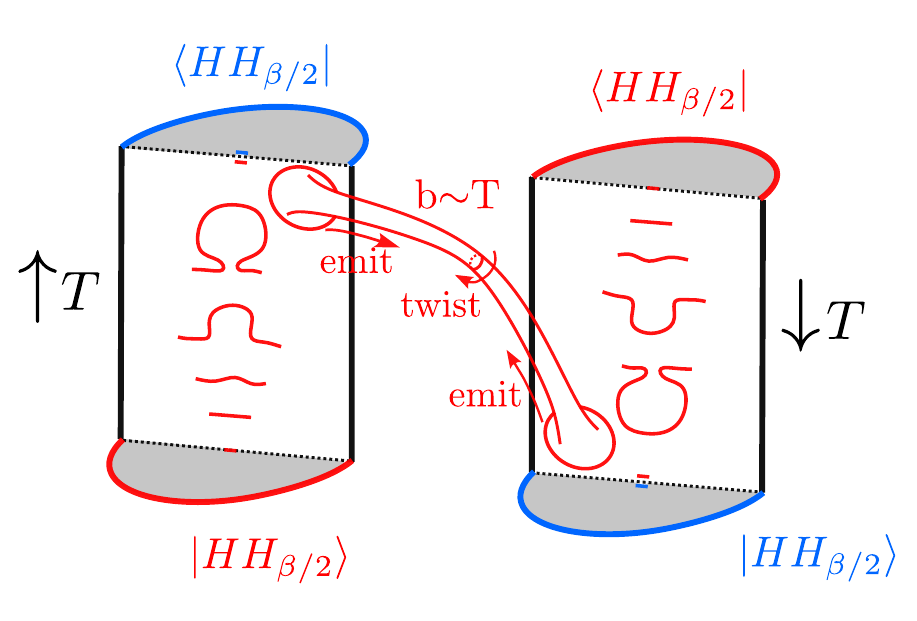}} 
    \subfigure[]{\includegraphics[width=0.4\textwidth]{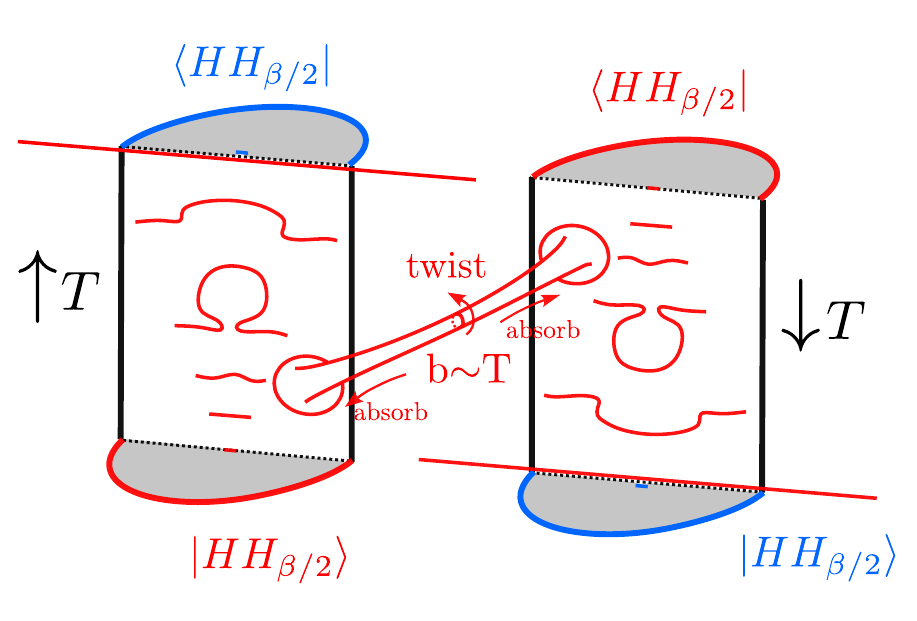}} 
    \caption{(a) \textit{Emitting process}: The length of ERB is shortened by the emission of a baby universe with size T, resulting a non-decaying return amplitude. The twist factor gives linear ramp. (b) \textit{Absorbing process}: The  ERB length is enlarged through the absorption of a baby universe, which further decreases the return amplitude. Absorption process provides sub-dominant contribution to SFF.}
    \label{trump_SSF}
\end{figure}

We note that since  the baby universe exchange originates from a Euclidean geometry, such process must be time-reversal symmetric in the Lorentzian picture. This means the baby universe can be emitted or absorbed with equal probability. However, since the absorbing process further increases the ERB length, it is always sub-dominant in the SFF (and two-point functions). Nevertheless, keeping a time-reversal symmetric treatment by including the absorption process is essential for the proper normalization of the total probability, a point we will come back to in Sec.\ref{grayhole_sec}.  We illustrate both emission and absorption process in the Lorentzian signature in Fig. \ref{trump_SSF}.       

The above picture accounts for the ramp at times earlier than  the Heisenberg time and receives only small corrections in the genus expansion. In the next section we extend this geometrical picture to later time, we will see that this results in a similar geometrical picture with an effective cutoff on the twist factor for large baby universes.

\vspace{10pt}
\section{FZZT-branes and twist factor cutoff}
\label{FZZT_sec}

\begin{figure}[htbp]
  \begin{center}
   \includegraphics[width=7cm]{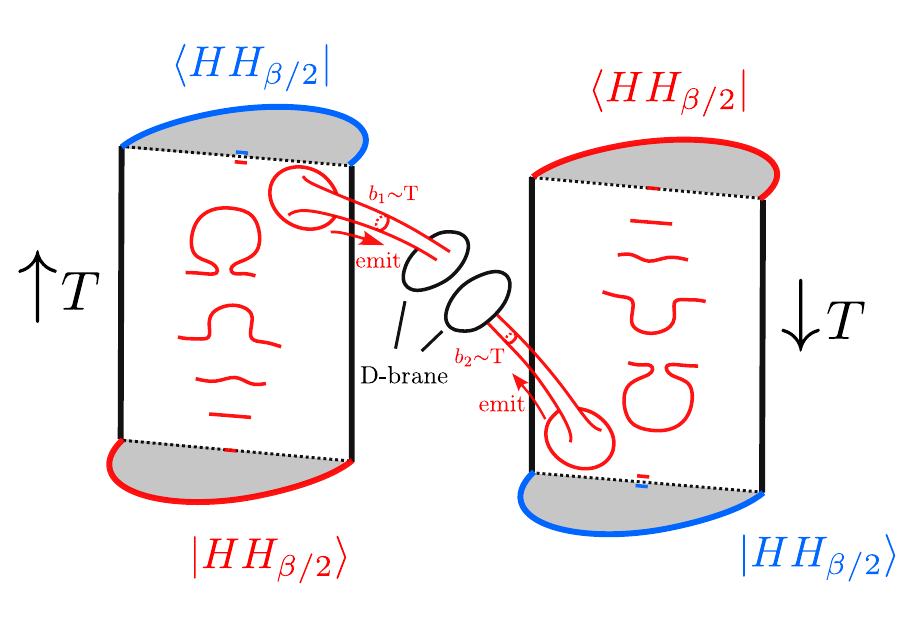}
  \end{center}   
\vspace*{-0.5cm}
\label{brane_SFF}
\caption{D-brane picture for plateau.}
\end{figure}

After the Heisenberg time non-perturbative effects dominate the SFF. In \cite{Saad:2019pqd} it was suggested that this regime could be described by including D-brane boundaries, such that emitted baby universes end on D-branes instead of being exchanged. This geometrical picture eliminates the twist factor measure, 
 resulting in a plateau in the SFF.  The Lorentzian picture is illustrated in Fig.\ref{brane_SFF}. In this section we substantiate this geometrical picture using FZZT or eigen-branes, and in the next section we  then apply it to the calculation of the firewall probability at late times and the gray hole conjecture.

\vspace{5pt}
\subsection{FZZT branes}

JT gravity is dual to a double scaled matrix integral \cite{Saad:2019lba}, i.e.  the topological expansion of JT gravity (\ref{muti_bdy_pati})  can be computed by the ensemble average of the matrix operators $\mathrm{Tr} e^{\beta_i M}$:
 \begin{align}
      \langle Z(\beta_1) ... Z(\beta_n) \rangle_{\mathrm{JT}}
      &=\sum_{g=0}^{\infty}e^{S_0 (2-2g-n)} Z_{g,n}(\beta_1,...,\beta_n),
      \\
     & =\frac{1}{Z_L}\int \mathcal{D}H \, e^{-L\, \mathrm{Tr}V(H)}  \, \mathrm{Tr} \, e^{\beta_i H} ... \, \mathrm{Tr} \, e^{\beta_n H},
     \label{dual_mat}
 \end{align}
where H is $L\times L$ ($L\gg 1$) Hermitian matrix with potential $V(H)$ and $Z_L$ is partition function for this matrix ensemble: $Z_L=\int \mathcal{D}H e^{-L\, \mathrm{Tr}V(H)} $.

As is common in the matrix/gravity correspondence, each eigenvalue of the matrix corresponds to a D-brane. On the gravity side, fixing an eigenvalue can be realized by insertion of a D-brane boundary with fixed energy and dilaton value \cite{Blommaert:2019wfy}, that is the DN boundary condition discussed in \cite{Goel:2020yxl}. In minimal string theory these corresponds to FZZT branes\footnote{JT gravity can be described as the p$\rightarrow \infty$ limit of the (2,p) minimal string, see \cite{Saad:2019lba}.}, and we use that term in our context as well. In the dual matrix integral a boundary with fixed energy $\lambda$ corresponds to an operator insertion:
 \begin{align}
    \mathcal{O}(\lambda)=-\int_{0}^{\infty}\frac{d\beta}{\beta}e^{\beta \lambda}Z(\beta) = \mathrm{Tr}\log(\lambda-H)-\frac{L}{2}V(\lambda),
    \label{FZZT-bdy}
 \end{align}
where $\frac{1}{\beta}$ means we consider an  unmarked boundary and divide the gauge group volume $\mathrm{Vol}(S^1)=\beta$. Then the FZZT brane operator is the exponential, corresponding to an insertion of any number of boundaries:
\begin{align}
    \psi (\lambda)=\exp \left( \mathcal{O}(\lambda) \right) =\det(\lambda-H)\exp\left( -\frac{L}{2}V(\lambda)\right).
    \label{FZZT-brane}
\end{align}

A complete picture of the correspondence can be obtained in the framework of Kodaira-Spencer theory \cite{Post:2022dfi, Altland:2022xqx} following the Dijkgraaf -Vafa conjecture \cite{Dijkgraaf:2002fc, Dijkgraaf:2002vw}. The eigenvalue branes correspond to flavor branes (see also \cite{Blommaert:2019wfy,Blommaert:2021fob}), which geometrize the eigenvalue decomposition of the matrix integral\footnote{The original matrix integral, which is the complete UV description, is visible in this framework as a collection "color branes" which undergo a geometric transition to provide the gravitational description.}. The full result as described by the matrix integral is obtained by summing over contributions from all possible configurations of the FZZT branes $\psi (\lambda_i), i=1,\cdots, L$. 

The description in terms of $L$ FZZT branes is fully non-geometrical and does not make immediate contact with the results of the genus expansion discussed above. Next we discuss fixing some number $1 \ll n \ll L$ eigenvalues, in order to retain the geometrical interpretation and the picture of "wormhole shortening" discussed above. This yields the picture suggested in \cite{Saad:2019pqd} of baby universes ending on branes.

\vspace{5pt}
\subsection{Partially fixed ensemble as an effective long time description}
Suppose we (temporarily) fix $n$ eigenvalues $\lambda_1$,$\lambda_2$,...,$\lambda_n$ in the partition function of the matrix ensemble $Z_L$ \cite{Blommaert:2019wfy}, which we denote as  $Z_L^{\lambda_1,\lambda_2,...,\lambda_n}$:
\begin{align}
    Z_L^{\lambda_1,...,\lambda_n}\equiv \int \prod_{i=n+1}^{L}  \left\{ d\lambda_i  \right\}  e^{-L  \sum_{j=1}^L V(\lambda_j)}\Delta(\lambda_1,...,\lambda_L), 
    \label{fix_partition}
\end{align}
where $\Delta(\lambda_1,...,\lambda_L)$ is the Vandermonde determinant that is responsible for  level repulsion. A nice property of Vandermonde determinant can be related with FZZT brane operators as follows:
\begin{equation}
    e^{-L\sum_{i=1}^{L}V(\lambda_i)} \Delta(\lambda_1,...,\lambda_L) =\Delta(\lambda_1,...,\lambda_n)\left( \prod_{i=1}^n
     \psi^2(\lambda_i) \right)  e^{-L\sum_{j=n+1}^{L}V(\lambda_j)} \Delta(\lambda_{n+1},...,\lambda_L). 
     \label{determinant_brane}
\end{equation}
Here we assumed all the eigenvalues $\lambda_i$ are different and $L \gg n\gg 1$. Using this, (\ref{fix_partition}) can be cast into the form:
\begin{align}
    Z_L^{\lambda_1,...,\lambda_n} =Z_{L-n} \, \Delta(\lambda_1,...,\lambda_n)  \,  \langle \psi^2(\lambda_1)...\psi^2(\lambda_n) \rangle_{L-n}^{\lambda_1,...,\lambda_n}.
\label{partition_brane}
\end{align}

This shows that a partition function with $n$ fixed energy eigenvalues is equivalent to keeping $n$  FZZT branes at fixed energy while coarse-graining (tracing over) the remaining unfixed energy branes. The FZZT branes are now interpreted as boundary conditions for JT universes generated by averaging over the remaining $L-n$ eigenvalues\footnote{When $n \ll L$ the difference between averaging over $L$ and $L-n$ eigenvalues in non-perturbative}.

\begin{figure}[htbp]
  \begin{center}
   \includegraphics[width=14cm]{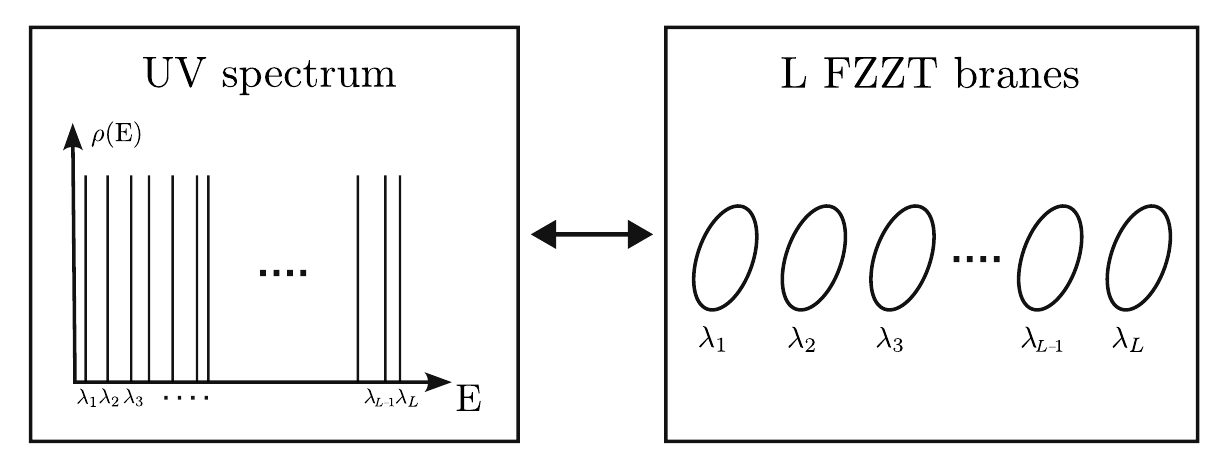}
  \end{center}   
\vspace*{-0.5cm}
\begin{center}
   \includegraphics[width=14cm]{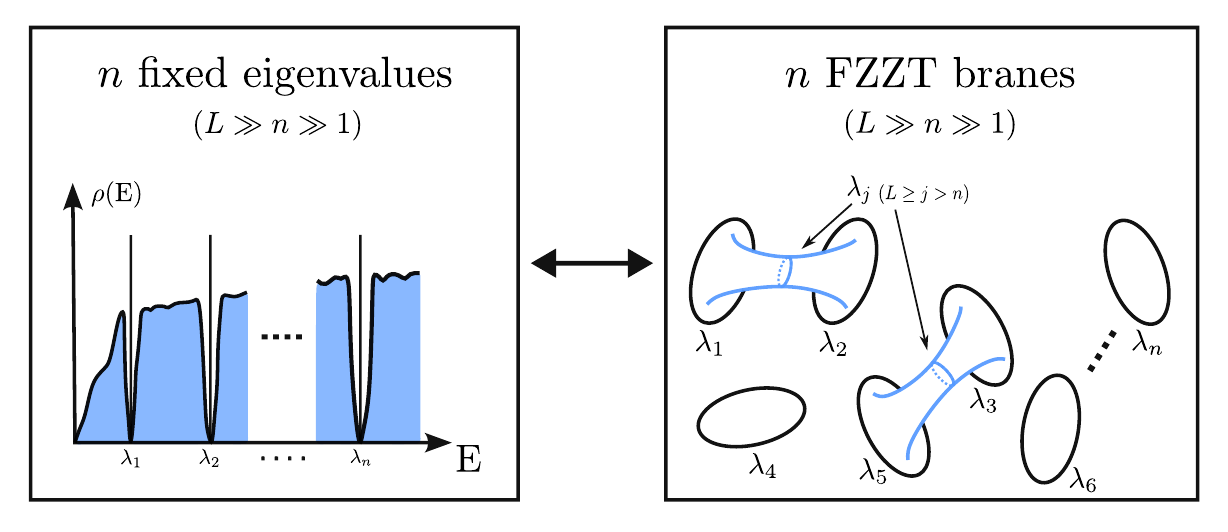}
  \end{center} 
  \vspace*{-0.5cm}
\caption{}
\label{UV-brane}
\end{figure}

When approaching $n=L$ where all energy levels are fixed,  the LHS of (\ref{partition_brane}) becomes a single member of the ensemble and the RHS is then $L$ FZZT branes without any averaging. This provides a description of a single member of the ensemble (with $L$ discrete energy levels) as a system of $L$  disconnected FZZT branes. We refer to the completely fixed situation as the microscopic or UV picture.
 The coarse-graining process pf $L-n$ branes allows the fixed FZZT branes to connect via Riemann surfaces, formed by the averaged branes, with the leading order contribution being disks and cylinders \cite{Blommaert:2019wfy}.  We illustrate this picture in Fig.\ref{UV-brane}.

\vspace{5pt}
 \subsection{Baby universes ending on branes}

From the gravity perspective, the description in terms of continuous spacetimes ending on D-branes is a hybrid of coarse-graining over most eigenvalues (resulting in a perturbative third quantized description, in other words the genus expansion) and singling out some eigenvalues for more precise non-perturbative treatment. This is natural when constructing an effective field theory for quantities that probe the physics at the scale of the level spacings, for example the exchange of large baby universes or the firewall probability. When discussing such quantities it is useful to integrate out the fast modes resulting in an effective theory for the slow modes. This identifies the $n$ fixed branes as eigenvalues in a single microcanonical window, in close vicinity of each other and the energies probed. The rest of the eigenvalues are "fast modes" that should be integrated out. In the time scales we are interested in they are always perturbative and thus integrating them out can be done in the genus expansion. This is similar to a Wilsonian picture in which we form an effective action for the slow modes (which are still to be integrated over to obtain a fully averaged result).

\begin{figure}[htbp]
  \begin{center}
   \includegraphics[width=7cm]{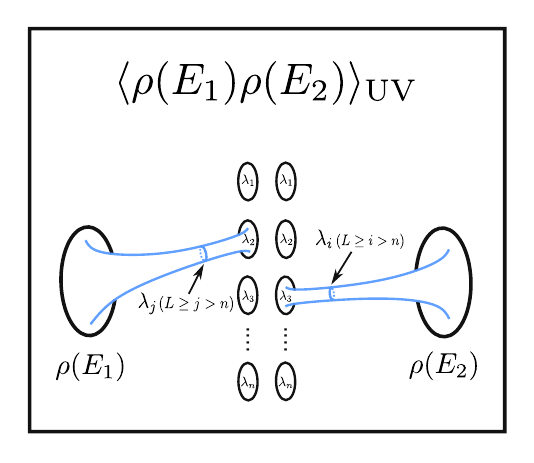}
  \end{center}   
\vspace*{-0.5cm}
\caption{}
\label{UV_SFF}
\end{figure}

For example for the late time spectral form factor \cite{Blommaert:2019wfy,Blommaert:2020seb}, we can use (\ref{determinant_brane}); trading the partition function for the  density of states, we obtain:
\begin{align}
\langle \rho(E_1) ... \rho(E_k) \rangle_L^{\lambda_1,...,\lambda_n} &= \frac{\langle \rho(E_1) ... \rho(E_k) \rho(\lambda_1) ...\rho(\lambda_n) \rangle_L}{\langle \rho(\lambda_1) ...\rho(\lambda_n) \rangle_L},
    \label{rho-rho} \\
      \langle \rho(\lambda_1)... \rho(\lambda_n) \rangle_L &=\frac{1}{(2\pi)^n} \Delta(\lambda_1,...,\lambda_n) \langle \psi^2(\lambda_1) ...\psi^2(\lambda_n) \rangle_{L-n}^{\lambda_1,...,\lambda_n},
    \label{rho-brane}
\end{align}
where we still set all $\lambda_i$ to be distinct and $L\gg n\gg k\geq1$ and $L-n\gg k$.  For the two-point function of the density of states the fully fixed expression $\langle \rho(E_1) \rho(E_2)\rangle_{\mathrm{UV}}$ corresponds to taking $k=2$ and setting  $n=L$. As a result, $\langle \rho(E_1) \rho(E_2)\rangle_{\mathrm{UV}}$ will be totally discrete and factorized \cite{Blommaert:2019wfy}:
\begin{align}
\langle \rho(E_1) \rho(E_2)\rangle_{\mathrm{UV}} &\equiv
    \langle \rho(E_1) \rho(E_2)\rangle_L^{\lambda_1,...,\lambda_n}, \nonumber
    \\
    &\approx \sum_{i=1}^n \delta(E_1-\lambda_i) \sum_{j=1}^n 
    \delta(E_2-\lambda_j),\nonumber
    \\
    &=\langle \rho(E_1)\rangle_L^{\lambda_1,...,\lambda_n} \langle \rho(E_2)\rangle_L^{\lambda_1,...,\lambda_n}.
\end{align}
As is mentioned in \cite{Blommaert:2019wfy}, this amounts to introducing a complete set of baby universes to achieve factorization. This is also similar to the half-wormhole discussed e.g. in \cite{Saad:2021rcu,Saad:2021uzi,Blommaert:2021fob}. This UV picture is presented in Fig.\ref{UV_SFF}.


Now we move to the effective description defined above, averaging over $L-n$ eigenvalues outside of a fixed microcanonical window, and discuss the exchange of large baby universes in that description.
A baby universe of fixed geodesic length $b$ corresponds to insertion of a geodesic boundary with zero extrinsic curvature and fixed boundary metric, which is referred as Neumann-Dirichlet (ND) boundary condition in \cite{Goel:2020yxl}. This operator can be obtained by inverse Laplace transform of an FZZT boundary in (\ref{FZZT-bdy}) \cite{Blommaert:2021fob,Goel:2020yxl}:
\begin{align}
    Z_{B}(b)&=-\frac{1}{2\pi i}\int_{\mathcal{C}} dz \, e^{bz} \, \mathcal{O}(-z^2)= \nonumber
    \\
    &=\frac{2}{b} \mathrm{Tr}\cos \left( b H^{1/2} \right) - \frac{2}{b}\int_0^{\infty} dE \, \rho_0(E) \cos(bE^{1/2}).
\end{align}
where $\mathcal{C}$ is the inverse Laplace contour, $H$ is $n\times n$ matrix with eigenvalues corresponding to the fixed $\lambda_i$ ($i=1,...,n$), and $Z_B(b)$ is baby universe operator with geodesic length b. We can the write the double trumpet contribution to  SFF in terms of the baby universe operators:
\begin{align}
    |Z(\beta+iT)|_{\text{double trumpet}}^2 &= \int_0^{\infty}b_1 db_1 \int_0^{\infty}b_2 db_2 \, \psi_{\mathrm{Tr}}(\beta+iT,b_1)  \, \psi_{\mathrm{Tr}}(\beta-iT,b_2)  \, Z_B(b_1)Z_B(b_2).
    \label{hw_SFF}
\end{align}
In the  UV picture where n FZZT branes are inserted to span microcanonical window so that everything is discrete and factorized, (\ref{hw_SFF}) is just a multiplication of two half-wormholes.
To obtain our effective description we coarse grain over the remaining $n$
eigenvalues. The resulting gravity with branes calculation gives then
 \begin{align}
     \langle Z_B(b_1) Z_B(b_2)\rangle_n &= \frac{4}{b_1 b_2}\left(\langle \mathrm{Tr}\cos ( b_1 H^{1/2} )  \mathrm{Tr}\cos  (b_2 H^{1/2}) \rangle - \langle \mathrm{Tr}\cos ( b_1 H^{1/2}) \rangle \langle   \mathrm{Tr}\cos ( b_2 H^{1/2} )\rangle\right), \nonumber
     \\
     &=\frac{4}{b_1 b_2} \langle \mathrm{Tr}\cos ( b_1 H^{1/2} )  \mathrm{Tr}\cos  (b_2 H^{1/2}) \rangle_{\mathrm{connected}},\nonumber
     \\
     &=\frac{1}{b_1 b_2} \langle \mathrm{Tr}e^{(\pm) i b_1 H^{1/2}} \mathrm{Tr}e^{(\pm)^{\prime} i b_2 H^{1/2}} \rangle_{\mathrm{connected}},
     \label{two-point BU}
 \end{align}
where in the first line we used $ \langle \mathrm{Tr}\cos ( b_1 H^{1/2}) \rangle \approx \int_0^{\infty} dE \rho_0(E) \cos (bE^{1/2})$, in the second line $\langle ... \rangle_{\mathrm{connected}}$ means taking the connected part,  and in the third line we used the shorthand notation $e^{(\pm)ix}=e^{ix}+e^{-ix}$. Note that (\ref{two-point BU}) is similar to the full SFF with $b$ conjugate to $H^{1/2}$ (similar to $T$ being conjugate to $H$). As we see below this similarity is manifested in a structure similar to the ramp-plateau transition. We emphasize that $H$ is now the Hamiltonian in the microcanonical window, with eigenvalues $\lambda_1, \cdots, \lambda_n$, realized geometrically as a configuration of D-branes JT universes can end on.

When $b$ becomes large the expression above is sensitive to small energy differences, i.e. we should use our effective description for a small microcanonical window. Expanding in the energy basis,
\begin{align}
     \langle Z_B(b_1) Z_B(b_2)\rangle_n &= \frac{1}{b_1 b_2} \int_0^{\infty} dE_1 dE_2 \, e^{(\pm) i b_1 E_1^{1/2}+(\pm)^{\prime} i b_2 E_2^{1/2}} \langle \rho_0(E_1) \rho_0(E_2)\rangle_{\mathrm{con}}.
\end{align}
and changing variables: 
\begin{equation}
    E= \frac{E_1+E_2}{2},\quad  \omega= E_1-E_2,
\end{equation}
The two point function of baby universe operators becomes for $E \gg 1$ \footnote{More precisely, the limit we are taking is: $E\gg1$, $\Delta E\ll1$ and $\rho(E)\gg1$ such that $\Delta E\rho(E)\gg1$ so that this small window still have large numbers of states,  and $\sqrt{E} b_1, \sqrt{E}b_2\gg 1$ so that the diagonal approximation produces $\delta(b_1-b_2)$.  $1/\rho(E)$ represents the level spacing of this microcanonical window.}
\begin{align}
    \langle Z_B(b_1) Z_B(b_2)\rangle_n &\approx \frac{e^{i\sqrt{E}[(\pm)b_1+(\pm)^{\prime}b_2]}}{b_1b_2}\int_{-\infty}^{\infty}d\omega e^{i\frac{\omega}{4\sqrt{E}}[(\pm)b_1 -(\pm)^{\prime}b_2]}\left( \delta(\omega)\rho_0(E)-\frac{\sin[\pi \rho_0(E)\omega]^2}{\pi^2 \omega ^2} \right), \nonumber
    \\
    &\approx\frac{2\pi\sqrt{E}\delta(b_1-b_2)}{b_1 b_2} \int_{-\infty}^{\infty} d\omega e^{(\pm)i\frac{\omega}{2\sqrt{E}}b_1} \left( \delta(\omega)\rho_0(E)-\frac{\sin[\pi \rho_0(E)\omega]^2}{\pi^2 \omega ^2} \right), \nonumber
    \\
    &=\frac{\delta(b_1-b_2)}{b_1b_2} \mathrm{min}\left\{b_1,4\pi \sqrt{E} \rho_0(E)\right\},
    \label{effective_twist}
\end{align}
In the first step of \ref{effective_twist} we see that for a microcanonical window of fixed high energy, where  we implicitly smear out $E$ over a small energy window so that the  oscillating phase at large $E$ generates the delta function, we project onto equal length wormholes $b_1=b_2=b$. 
\begin{equation}
    e^{i\sqrt{E}[(\pm)b_1+(\pm)^{\prime}b_2]} \approx \int_{E-\Delta E}^{E+\Delta E} d\tilde{E}  e^{i\sqrt{\tilde{E} }[(\pm)b_1+(\pm)^{\prime}b_2]} \approx 2\pi \sqrt{E} \delta\left((\pm)b_1+(\pm)^{\prime}b_2\right).
\end{equation}
The delta-function then select out $(\pm)=-(\pm)^{\prime}$, since $b_1$ and $b_2$ are both positive.

Within this microcanonical window the regime of large $b$ cannot be treated in the genus expansion, therefore we need to treat those eigenvalues more precisely as FZZT branes. We can then used the results from  random matrix theory for the $n \times n$ matrix $H$  \cite{Haake:2010fgh,Meh2004}:
\begin{align}
    \langle \rho_0(E_1) \rho_0(E_2)\rangle_{\mathrm{con}}= \delta (E_1-E_2)\rho_0(E_1) -\frac{\sin[\pi\rho_0(E_2)(E_1-E_2)]^2}{\pi^2 (E_1-E_2)^2},
\end{align}  
which completes the derivation of (\ref{effective_twist}). As anticipated (\ref{effective_twist}) exhibits a structure similar to the ramp-plateau of the SFF, we should not be surprised since $b$ plays a similar role to the time  $T$ in the SFF. We see that the more precise treatment of small range of eigenvalues as FZZT branes within the gravity description results in a simple prescription, where we use the handle-disk geometry but include a cutoff on the twist factor.

\vspace{5pt}
\subsection{Twist factor cutoff}
\label{twist cutoff}

\begin{figure}[htbp]
  \begin{center}
  \hspace{-2.2cm}
   \includegraphics[width=6cm]{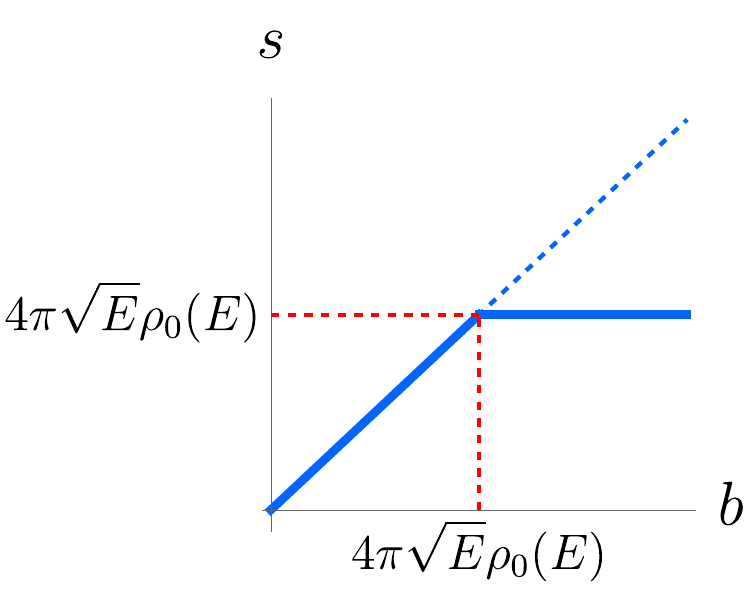}
  \end{center}   
\vspace*{-0.5cm}
\caption{Twist factor cutoff.}
\label{twist}
\end{figure}

Our last expression for $ \langle Z_B(b_1) Z_B(b_2)\rangle_n$ results  effectively in the following prescription within the gravitational description:

\newtheorem*{ansatz}{Twist Factor Cutoff}
\begin{ansatz} 
   The twist factor $s$ of a baby universe is bounded by a cutoff of $4\pi\sqrt{E}\rho_0(E)$.
   \begin{itemize}
       \item When the size of a  baby universes $b$ is smaller than $4\pi\sqrt{E}\rho_0(E)$, the baby universe will be traded between the systems and form a wormhole. This results in a connected Riemann surface with a usual twist factor equals to its size $b$.
       \item When the size of a baby universe is larger than $4\pi\sqrt{E}\rho_0(E)$, this baby universes will end on one of the $n$ D-branes in a microcanonical window. The effect of accounting for these non-perturbative D-brane effects is summarized by replacing the twist factor by $4\pi \sqrt{E} \rho_0(E)$.
       \item We draw this cutoff prescription in Fig.\ref{twist}.
   \end{itemize} 
\end{ansatz}

By construction this cutoff reproduces the ramp-plateau transition of the SFF by allowing the baby universe to end on FZZT branes, as suggested in \cite{Saad:2019pqd}. Indeed, taking (\ref{effective_twist}) back to (\ref{hw_SFF}), we find:
\begin{align}
    \langle  |Z(\beta+iT)|^2 \rangle_{\text{double trumpet}} =  \int_0^{\infty} db  \mathrm{~min}\left\{b,4\pi \sqrt{E} \rho_0(E)\right\} \, \psi_{\mathrm{Tr}}(\beta+iT,b)  \, \psi_{\mathrm{Tr}}(\beta-iT,b).
    \label{cutoff_twist}
\end{align}
Comparing (\ref{cutoff_twist}) with the genus expansion of JT gravity in (\ref{muti_bdy_pati}), we observe that the original twist factor $b$ is replaced by $\mathrm{min}\left\{b,4\pi \sqrt{E} \rho_0(E)\right\}$. This implies that when the baby universe is smaller than $4\pi \sqrt{E} \rho_0(E)$, we can coarse-graining all D-branes and reproduce the wormhole construction.  However, when the size of baby universe exceeds $4\pi \sqrt{E} \rho_0(E)$, we can no longer coarse-grain the entire spectrum and need to leave some branes for more precise treatment as D-branes in JT gravity. This results in the twist factor being  replaced by $4\pi \sqrt{E} \rho_0(E)$, no longer scaling with the size of baby universe.

\begin{figure}[htbp]
  \begin{center}
   \includegraphics[width=15cm]{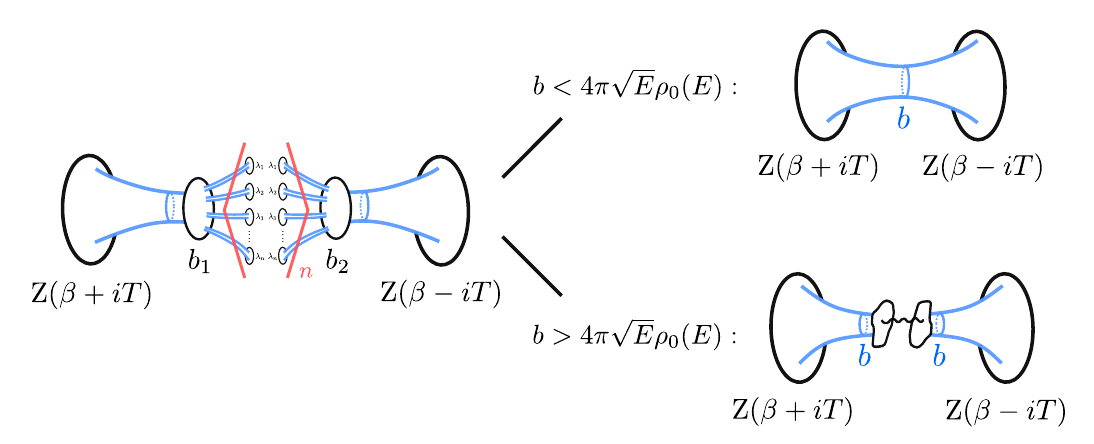}
  \end{center}   
\vspace*{-0.5cm}
\caption{Emergent ramp-plateau structure when accounting for the $n$ FZZT branes. For $b<4\pi \sqrt{E} \rho_0(E)$, the fixed $n$ branes can be coarse-grained together with the $L-n$ branes, and the baby universe form a wormhole, thus reproducing the ramp. For $b>4\pi \sqrt{E} \rho_0(E)$, baby universes end on FZZT branes which when integrated over the plateau. This could be visualized as the baby universe ending on effective correlated branes as drawn. Note that this ramp-plateau structure here pertains to the size of baby universes,  not time.}
\label{cutoff_SFF}
\end{figure}

We can use the twist factor cutoff prescription to repeat Saad's wormhole story of the SFF (or correlations functions). At early time, the disk contributions dominate and  gives the slope, as no baby universe are exchanged. At later times, the double-trumpet dominates the SFF, allowing the two system to emit or absorb baby universes of any size. The only way to provide a non-decaying return amplitude is by emitting a baby universe of size $T$. For early enough times the baby universes traded are small compared to the twist factor cutoff. Such relatively small baby universes are traded between the two systems and thus include a twist factor is proportional to $T$, giving rise to the ramp. At late enough time, beyond the Heisenberg time, large baby universes take over the job of stopping the decay of the return amplitude. These large baby universe are not able to be traded but rather end on D-branes. Utilizing the resulting twist factor cutoff, they will  provide a non-decaying amplitude without extra twist factors to bring in time dependence, recovering the plateau at times beyond the  Heisenberg time.

\subsection{Effective brane description of the plateau}

Regardless of the size of the baby universe, the coarse graining over most of the spectrum leads to non-factorized structure, visible in  the delta-function enforcing $b_1=b_2$. This can be visually represented as in figure \ref{cutoff_SFF}, where for small $b$ the constraint $b_1=b_2$ is enforced by having a smooth surface, a wormhole. For large $b$ the non-factorization for is represented by correlated (or entangled) effective branes, similar in spirit to the suggestion in \cite{Saad:2019pqd}.

In this sense the coarse graining or disorder averaging leads to entanglement, effective branes that are in a Bell pair state. To see this, let us rewrite the SFF in UV picture in (\ref{hw_SFF}) by inner products of different states in Hilbert space:
\begin{align}
     |Z(\beta+iT)|_{\text{double trumpet}}^2 &= \Big{(} \langle HH_{\beta+iT}| \otimes \langle HH_{\beta-iT}| \Big{)} \Big{(} |\text{BU} \rangle_{H} \otimes |\text{BU} \rangle_{H} \Big{)},
\end{align}
where $|\text{BU} \rangle_{H}$ represents the baby universe state with inclusion of the FZZT brane contribution:
\begin{equation}
    |\text{BU} \rangle_{H}= \int_{0}^{\infty} b~ db ~Z_B(b) ~|b\rangle ,
\end{equation}
and $H$ denotes the dependence of $n\times n$ H matrix in $Z_B(b)$. In this UV regime, the two baby universes are not entangled; rather, they contribute as a simple product state, $|\text{BU} \rangle_{H} \otimes |\text{BU} \rangle_{H}$.

At IR, however, disorder averaging brings in delta-function, so that SFF now becomes:
\begin{align}
   \langle  |Z(\beta+iT)|^2 \rangle_{\text{double trumpet}}=  \Big{(} \langle HH_{\beta+iT}| \otimes \langle HH_{\beta-iT}| \Big{)} \Big{(} \int_0^{\infty} s(b) db ~|b\rangle \otimes |b\rangle \Big{)}.
   \label{entangle_BU}
\end{align}
Here, the two baby universes become highly entangled ( large $b$ corresponds to the maximal entanglement and produces plateau). We see that, in the IR, the transition from a factorized geometry to a non-factorized one corresponds to the emergence of entanglement between the two baby universe states. This aligns with the ER=EPR conjecture.

\vspace{10pt}
\section{Effective time slice and the gray hole conjecture}
\label{grayhole_sec}

The modified story of Saad's wormhole is particularly useful for studying quantities that require non-perturbative completion at all times.  It encodes non-perturbative effects coherently while still preserving a clear geometric picture.   In this section we calculate the two-point function in the handle-disk geometry, aiming for extracting the probability for the late time Cauchy slice being in a black or white hole state, confirming the "gray hole conjecture". We will see how the inclusion of non-perturbative effects through the above twist factor cutoff preserves a clear geometrical picture and ensures that the sum of probabilities is always one at all times.  This provides a unified derivation of the gray hole conjecture (for various probes used previously in the literature \cite{Stanford:2022fdt, Blommaert:2024ftn,Iliesiu:2024cnh,Miyaji:2024ity}) and extends that to new situations (e.g. with the inclusion matter fields).



\vspace{5pt}
\subsection{Handle-disk calculation}
To this end, let us decompose the two-point function on handle-disk geometry as follows:
\begin{align}
    G_{\Delta}(T)&=\sum_{\gamma_i=\{\gamma_1,\gamma_2,\gamma_3\}} \adjincludegraphics[width=4cm,valign=c]{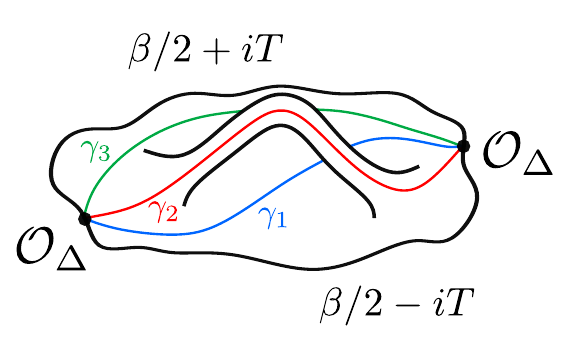},
    \nonumber \\
    &\approx \int_{\mathcal{F}(b_1)} d b_1 d\tau_1 \adjincludegraphics[width=4cm,valign=c]{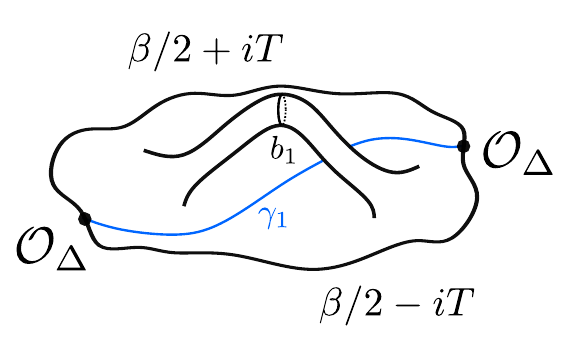} +\int_{\mathcal{F}(b_2)} d b_2 d\tau_2 \adjincludegraphics[width=4cm,valign=c]{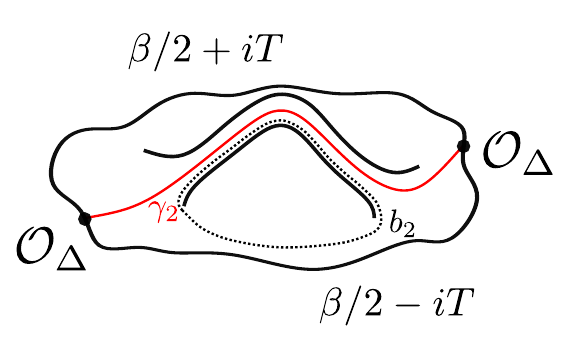},
    \nonumber \\
    & = \int_{0}^{\infty} s_1~ db_1 \adjincludegraphics[width=4cm,valign=c]{figure/handle-disk2.pdf}.
\end{align}
The gravity calculation of the two-point function requires summing over all inequivalent geodesics connecting the boundary operators.  Thus in the first line we summed over three classes of geodesics in the bulk geometry. Among these, $\gamma_3$ is unaffected by the presence the handle and remains identical to the disk contribution.\footnote{The handle merely acts as a renormalization of disk area for $\gamma_1$.}. Since our interest is late-time behavior of this two-point function while the disk contribution decays, we keep only the $\gamma_1$ and $\gamma_2$ contributions in the second line. Importantly, to avoid over-counting, the Weil-Petersson measure $b_i\wedge \tau_i$ for the bulk moduli should be integrated over the fundamental domains of the mapping class group $\mathcal{F}(b_i)$.  However, cutting along $\gamma_1$ and $\gamma_2$ in each figure yields the same geometry, a two-sided trumpet. Therefore, summing over different geodesics can be achieved instead by integrating over two fundamental domains while  fixing the  geodesic, as explained in \cite{Saad:2019pqd}. This simplifies the calculation since $\sum\limits_{i=1}^2 \int_{\mathcal{F}(b_i)} db_i d\tau_i=\int_{0}^{\infty} db \int_0^{s} d\tau$, where $s$ is the twist factor. Consequently, in the third line, we can just utilize a single geometry without mapping class group constraints.

Cutting the handle-disk along the geodesic $\gamma_1$ results in two-sided trumpet with mixed boundaries:
\begin{align}   
    G_{\Delta}(\tau) \supset& \frac{e^{-S_0}}{Z(\beta)}  \int_{-\infty}^{\infty}d\ell \int_{0}^{\infty} s~ db  \adjincludegraphics[width=4cm,valign=c]{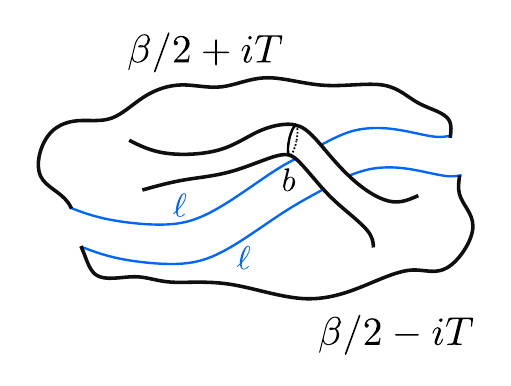 }  e^{-\ell\Delta}
    \label{2_p_trump}, \nonumber \\
    =&  \frac{e^{-S_0}}{Z(\beta)}   \int_{-\infty}^{\infty}d\ell \int_{0}^{\infty} s~ db   ~\psi_{\mathrm{Tr},\beta/2-iT}(\ell,b) ~\psi_{\mathrm{Tr},\beta/2+iT}(\ell,b) e^{-\ell \Delta}, \\
  =&   \frac{e^{-S_0}}{Z(\beta)}   \int_{-\infty}^{\infty}d\ell \int_{0}^{\infty} s~ db  \int_0^{\infty} dE_1 dE_2 \frac{\mathrm{cos}(b \sqrt{E_1}) \mathrm{cos}(b \sqrt{E_2})}{4\pi^2 \sqrt{E_1 E_2}}  \nonumber \\
  & \times e^{- (E_1+E_2)\beta/2 -i (E_1-E_2)T} \psi_{E_1}(\ell) \psi_{E_2}(\ell) ~ e^{-\ell \Delta}.
    \label{2_p_loop}
\end{align}
Under the change of variables $E= \frac{E_1+E_2}{2}$, $\omega =E_1-E_2$ and considering a small micro-canonical window with large fixed total energy $E \gg 1$, the two point function simplifies to:
\begin{align}
     G_{\Delta}(T)  \supset &  e^{-S_0}  \int_{-\infty}^{\infty}d\ell \int_{0}^{\infty} s~ db \int_{-\infty}^{\infty} d\omega \frac{\mathrm{cos}(b \sqrt{E+\frac{\omega}{2}}) \mathrm{cos}(b \sqrt{E-\frac{\omega}{2}})}{4\pi^2 E \rho_0(E)}\nonumber
     \\
    &\times e^{-i\omega T} \psi_{E+\frac{\omega}{2}}(\ell) \psi_{E-\frac{\omega}{2}}(\ell) e^{-\ell \Delta},
      \label{2-pt-before appro}
      \\
      \approx &   \int_{-\infty}^{\infty}d\ell \int_{0}^{\infty} s~ db  \int_{-\infty}^{\infty} d\omega \frac{\mathrm{cos}(b \sqrt{E+\frac{\omega}{2}}) \mathrm{cos}(b \sqrt{E-\frac{\omega}{2}})}{4\pi^2 E \rho_0(E)}  
      \nonumber \\
     & \times e^{-i\omega T} \Theta \left(log(E)+\ell \right) \frac{\mathrm{cos}\left( \frac{\omega}{E^{1/2}}\mathrm{arccossh}(e^{\ell/2}E^{1/2})\right)}{2\pi \rho_0(E)(E-e^{-\ell/2})^{1/2}} e^{-\ell \Delta},
     \nonumber \\
      =&  \int_{-\infty}^{\infty}dT_{\mathrm{eff}} \int_{0}^{\infty} s~ db \int_{-\infty}^{\infty} d\omega \frac{\mathrm{cos}(b \sqrt{E+\frac{\omega}{2}}) \mathrm{cos}(b \sqrt{E-\frac{\omega}{2}})}{2^2\pi^3 E \rho_0^2(E)}  \nonumber \\
      & \times e^{-i\omega T} \Theta (T_{\mathrm{eff}}) \, \mathrm{cos}(\omega T_{\mathrm{eff}}) \, e^{-\ell_{T_{\mathrm{eff}}} \Delta},
     \label{G-cos}
\end{align}
where in the second line we used the semiclassical approximation of $ \psi_{E+\frac{\omega}{2}}(\ell) \psi_{E-\frac{\omega}{2}}(\ell)$ given in \cite{Blommaert:2024ftn}, for convenience we include the full  derivation in Appendix \ref{appen_semi}. In the last line we define the effective time \cite{Blommaert:2024ftn} $T_{\mathrm{eff}}=\frac{1}{E^{1/2}}\mathrm{arccosh}(e^{\ell/2}E^{1/2})$. At the large $E$ limit, this transformation matches to Stanford and Yang's result \cite{Yang:2018gdb}: $\ell_{T_{\mathrm{eff}}}=2|E^{1/2}T_{\mathrm{eff}}|-\ln 4E$.

An important observation is that both $T_{\mathrm{eff}}$ and $b$ appear inside the cosine function. By expressing the cosine as sum of  exponentials, they will both have + branch and - branch. However, $T_{\mathrm{eff}}$ has an additional $\Theta(T_{\mathrm{eff}})$ function, which combines these two branches into a single branch. This ensures that, for a given analytical continuation of T ($>0$), we will not simultaneously obtain a black hole slice ($T_{\mathrm{eff}}>0$) and a white hole slice ($T_{\mathrm{eff}}<0$), unless a baby universe is emitted.

Next we again use the shorthand notation $\mathrm{cos}(x)\mathrm{cos}(y)=\frac{1}{4 }e^{(\pm)ix}e^{(\pm)^{\prime}iy}$ where $e^{(\pm)ix}=e^{ix}+e^{-ix}$. Then (\ref{G-cos}) is cast into the form:
\begin{align}
    G_{\Delta}(T)  \supset &   \int_{-\infty}^{\infty}dT_{\mathrm{eff}} \int_{0}^{\infty} s~ db  \frac{1}{2^4\pi^2 E \rho_0^2(E)} \int_{-\infty}^{\infty} \frac{d\omega}{2\pi} e^{i\omega (T_{\mathrm{eff}}-T)}\nonumber \\
   & \times e^{i\sqrt{E}b\left( (\pm)+(\pm)^{\prime}\right)+i\frac{\omega}{4\sqrt{E}}b \left( (\pm)-(\pm)^{\prime}\right)} e^{-\ell_{T_{\mathrm{eff}}} \Delta}.
\end{align}
If $ (\pm)=(\pm)^{\prime}$, the $e^{\pm i2\sqrt{E}b}$ term won't survive under the limit $E\gg1$ and the integration of $b$, thus we are left with $(\pm)=-(\pm)^{\prime}$, giving
\begin{align}
    G_{\Delta}(T)  \supset &   \int_{-\infty}^{\infty}dT_{\mathrm{eff}}~  \frac{1}{2^4\pi^2 E \rho_0^2(E)} \int_{0}^{\infty} s~ db \int_{-\infty}^{\infty} \frac{d\omega}{2\pi} e^{i\omega (T_{\mathrm{eff}}-T\pm\frac{1}{2\sqrt{E}}b)} ~e^{-\ell_{T_{\mathrm{eff}}} \Delta} ,
   \nonumber \\
    =&   \int_{-\infty}^{\infty}dT_{\mathrm{eff}}  ~\frac{1}{2^4\pi^2 E \rho_0^2(E)} \int_{0}^{\infty} s~ db ~
    \delta(T_{\mathrm{eff}}-T\pm\frac{1}{2\sqrt{E}}b) ~e^{-\ell_{T_{\mathrm{eff}}} \Delta} ,
    \nonumber \\
     =&   \int_{-\infty}^{\infty}dT_{\mathrm{eff}}~  \frac{1}{2 \sqrt{E} T_H^2} ~s(b)\Big{|}_{b=\pm 2\sqrt{E}(T_{\mathrm{eff}}-T)}
    ~e^{-\ell_{T_{\mathrm{eff}}} \Delta},
    \label{2pt_final}
\end{align}
 where we identified the Heisenberg time as $T_H=2\pi \rho_0(E)$. Note that in the third line we replace now $s \rightarrow s(b)$ so that we can now use our twist factor cutoff in Fig.\ref{twist} to incorporate the late time D-brane effects. 

 \vspace{5pt}
 \subsection{Incorporating the twist factor cutoff}
 
 In the final expression (\ref{2pt_final}), the twist factor $s(b)$ plays a role analogous to the conversion factor $\mathcal{F}(T_{\mathrm{eff}}|T)$ defined in \cite{Blommaert:2024ftn}\footnote{The expressions become identical when aligning the notation.}. For us the  conversion factor acquires a very geometric interpretation -- the two terms in (\ref{2pt_final}) represent the possibilities of emission or absorption of a baby universe: 
\begin{itemize}
    \item \textbf{Emitting Branch}: $T_{\mathrm{eff}}=T -\frac{1}{2\sqrt{E}}b$. 

     The Cauchy slice of two-sided black hole emits a baby universe of size $\frac{1}{2\sqrt{E}}b$, causing the effective time of this slice to be reduced. If $\frac{1}{2\sqrt{E}}b > T$, this emitting process will convert a black hole into a white hole. 
    
    \item \textbf{Absorbing Branch}: $T_{\mathrm{eff}}=T +\frac{1}{2\sqrt{E}}b$.
    
    The Cauchy slice of two-sided black hole absorbs a baby universe of size $\frac{1}{2\sqrt{E}}b$, causing the effective time of this slice to increase. This is a time reversed process of emitting a baby universe. This absorbing process will not turn a black hole into a white hole.
\end{itemize}

\begin{figure}[htbp]
  \begin{center}
   \includegraphics[width=6cm]{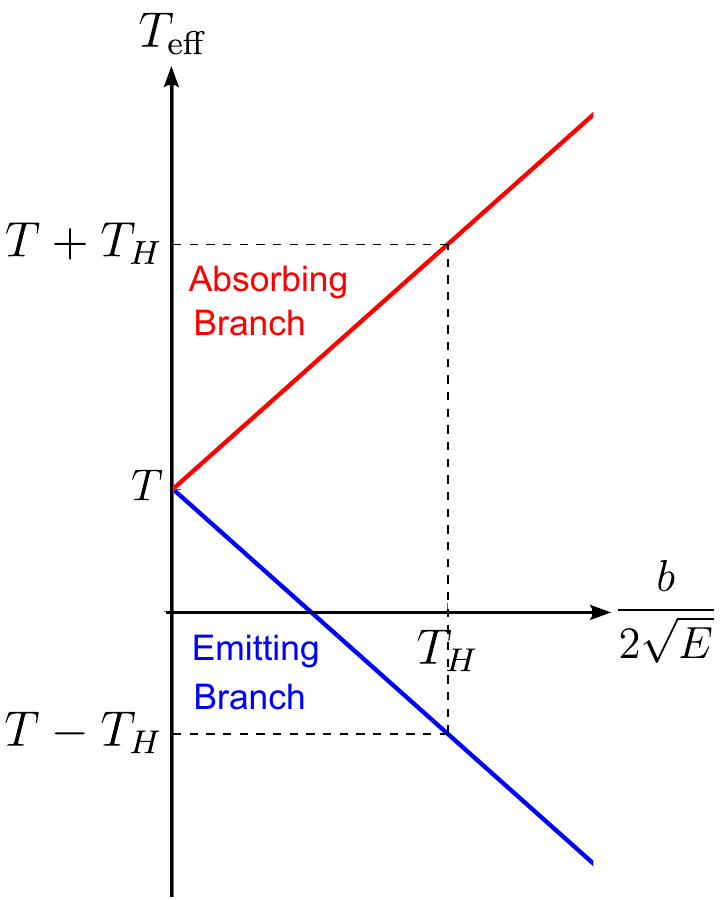}
  \end{center}   
\vspace*{-0.5cm}
\caption{Absorbing Branch (Red) and Emitting Branch (Blue). The black dashed lines represent the twist factor cutoff.}
\label{b-T}
\end{figure}

\begin{figure}[htbp]
\begin{tabular}{cc}
    \begin{minipage}{0.5\hsize}
     \begin{center}
     \includegraphics[width=7.4cm]{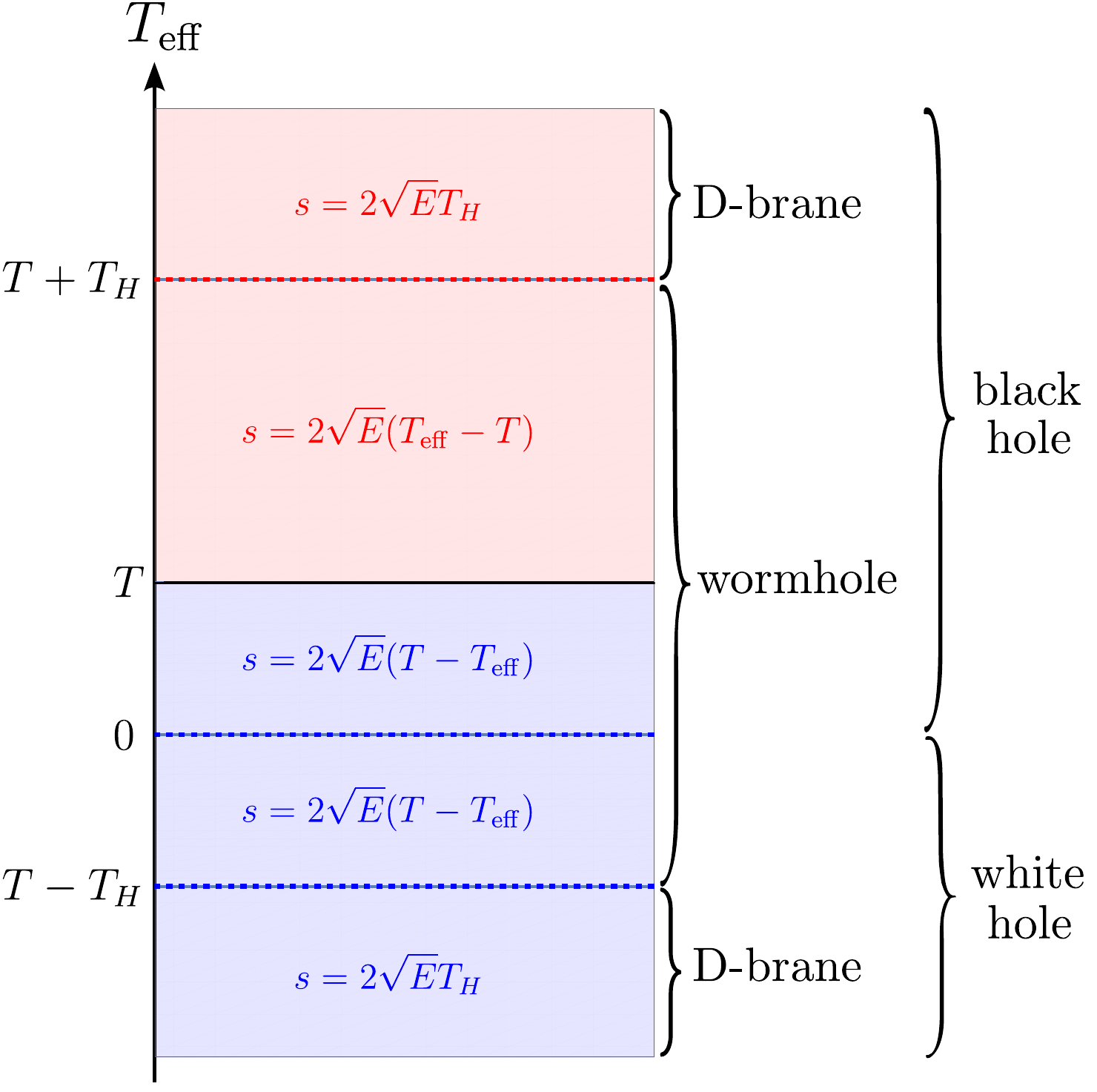}
     \subfigure{(a)}
    \end{center}
\end{minipage}
 \begin{minipage}{0.5\hsize}
     \begin{center}
     \includegraphics[width=7.4cm]{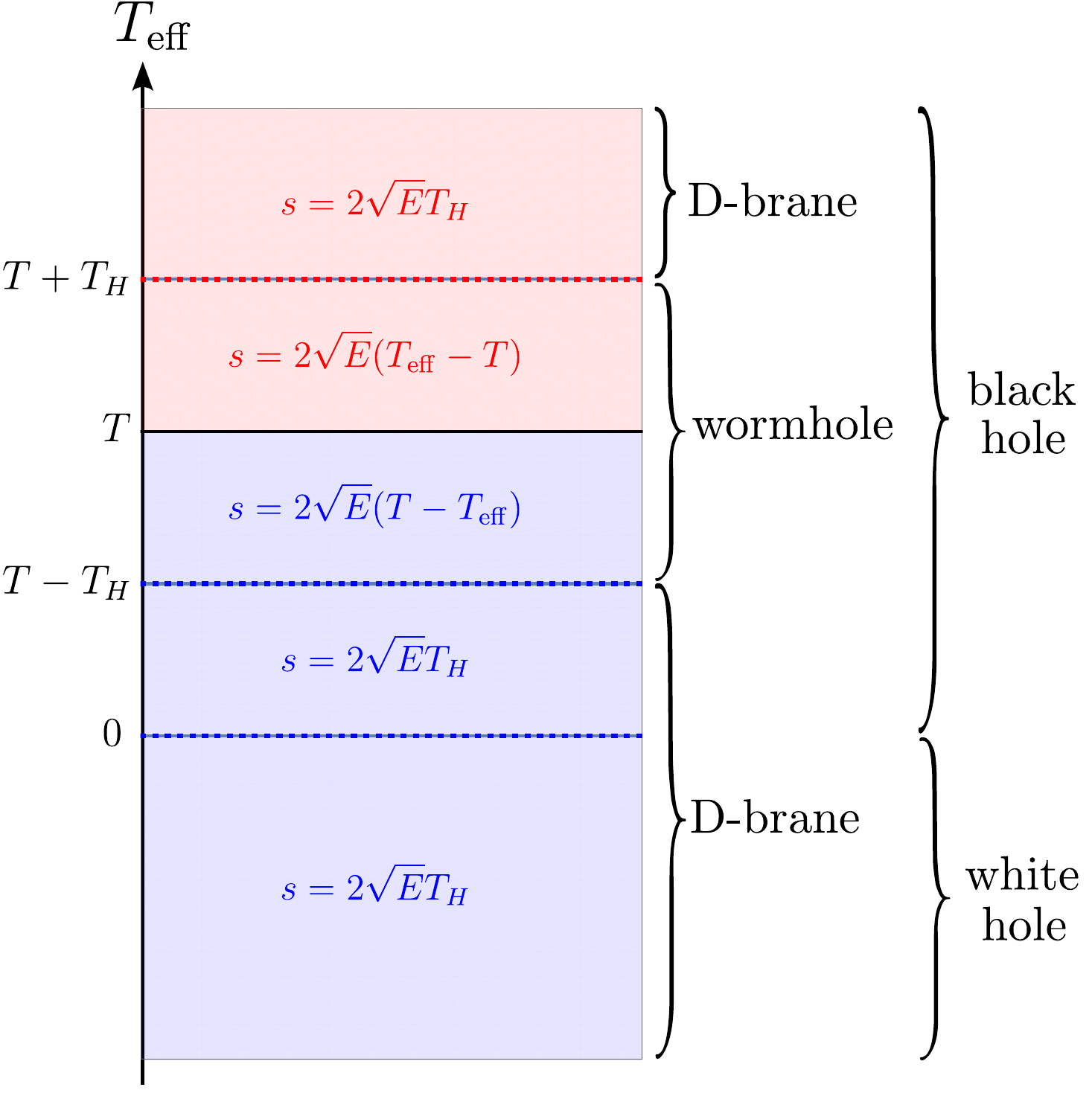}
     \subfigure{(b)}
    \end{center}
\end{minipage}
\end{tabular}
\caption{Twist factor at different effective times  with the red region the absorbing branch and the blue region the emitting branch.  (a): $T<T_H$. (b): $T>T_H$.}
\label{T-S}
\end{figure}

With our prescription the twist factor $s(b)$ is bounded by $2\sqrt{E}T_H$ when $b > 2\sqrt{E}T_H$. The effective time $T_{\mathrm{eff}}$ involves two possible sign choices for $b$, thus the relation between twist factor and $T_{\mathrm{eff}}$ becomes a little involved. We show the relationship between the effective time and the size of the baby universe in fig.\ref{b-T} and the corresponding twist factor in fig.\ref{T-S}. The various possibilities are explained below.

The probabilities for the late time slice being in a white hole state ($T_{\mathrm{eff}}<0$) or a black holes state ($T_{\mathrm{eff}} > 0$) can be extracted from (\ref{2pt_final}) by integrating over the prefactor of the disk two-point function $e^{-\ell_{T_{\mathrm{eff}}}\Delta}$. Below we compute them separately.
\vspace{5pt}
    \paragraph{White hole probability ($\mathcal{P}_{\mathrm{WH}}$)}

     The white hole probability arises solely from the emitting branch, with contributions depending on whether the emitted baby universe forms a  wormholes or ends on a D-brane.

\begin{itemize}
    \item For $T<T_{H}$, there are two cases (see Fig.\ref{T-S}(a)): 
    \begin{itemize}
        \item[(i)] $T-T_H <T_{\mathrm{eff}}<0$:
        
        The emitted baby universe forms a wormhole with twist factor  $s= 2\sqrt{E}(T-T_{\mathrm{eff}})$.
        \item[(ii)] $T_{\mathrm{eff}}<T-T_H$:
        
        The emitted baby universe ends on D-brane, with twist factor  $s=2\sqrt{E}T_H$.
    \end{itemize}
    The probability is then given by:
    \begin{align}
        \mathcal{P}_{\mathrm{WH}} (T) &=\frac{1}{T_H^2}\int_{T-T_H}^0 dT_{\mathrm{eff}} (T-T_{\mathrm{eff}}) + \frac{1}{T_H^2}\int_{-\infty}^{T-T_H} dT_{\mathrm{eff}} T_H, 
        \nonumber \\
        &=\frac{T}{T_H}-\frac{T^2}{2T_H^2}+\mathrm{const},
    \end{align}
     where "const" is a divergent constant:\footnote{This divergence arises from the infinite over-counting resulting from  an incomplete fixing of the of the mapping class group. Summing over geodesics in the bulk gets rid of S-transformation constraint (applied to the twist factor) on the handle. However, the remaining T-transformation sends $\ell\rightarrow\ell+nb$, where n is the winding number. Consequently, integrating both $\ell$ and b  to infinity over-counts infinitely many images under the T-transformation.}
\begin{align}
    \mathrm{const}=-\frac{1}{2}+\frac{1}{T_H}\int_{-\infty}^{0} dT_{\mathrm{eff}},
\end{align}
and needs to be regularized to zero by imposing $\mathcal{P}_{\mathrm{WH}}(T=0)=0$, as in \cite{Blommaert:2024ftn}.

    \item For $T>T_{H}$, 
    the only way to tunnel into a white hole is by emitting a large baby universe which then necessarily ends on a D-brane, with twist factor $s=2\sqrt{E}T_H$ (see Fig.\ref{T-S}(b)). The  corresponding probability is:
    \begin{align}
         \mathcal{P}_{\mathrm{WH}} (T) &=\frac{1}{T_H^2}\int_{-\infty}^0 dT_{\mathrm{eff}} T_H,
        \nonumber \\
        &=\frac{1}{2}+\mathrm{const}.
    \end{align}
\end{itemize}

\vspace{5pt}
\paragraph{Black hole probability ($\mathcal{P}_{\mathrm{BH}}$)}

 The  black hole probability is more complicated. First, we  need to include the disk contribution. Using the previous approximations of the  wavefunctions, the disk two-point function in a micro-canonical ensemble with $E\gg 1$ is:
\begin{align}
 G_{\Delta}(T)  \supset &  \frac{e^{-S_0}}{\rho_0(E)}  \int_{-\infty}^{\infty}d\ell \int_{0}^{\infty} sdb  \int_{-\infty}^{\infty} d\omega \rho_0(E+\frac{\omega}{2}) \rho_0(E-\frac{\omega}{2}) e^{-i\omega T} \psi_{E+\frac{\omega}{2}}(\ell) \psi_{E-\frac{\omega}{2}}(\ell) e^{-\ell \Delta} ,
 \nonumber \\
 \approx &    \int_{-\infty}^{\infty}dT_{\mathrm{eff}}  \delta(T_{\mathrm{eff}}-T)
    e^{-\ell_{T_{\mathrm{eff}}} \Delta}.
\end{align}
We see that the effective time for the disk is always $T$. 

Secondly, the handle-disk geometry now involves both emitting and absorbing branches. Taking these effects into account we can calculate the black hole probability for various times.
\begin{itemize}
\item For $ T < T_H $, there are three cases (see Fig.\ref{T-S}(a)):
\begin{itemize}
    \item[(i)] Emitting branch  ($0 < T_{\mathrm{eff}} < T$): 
    
    The emitted baby universe forms a wormhole, with twist factor $s= 2\sqrt{E}(T - T_{\mathrm{eff}})$.
    \item[(ii)] Absorbing branch ($T < T_{\mathrm{eff}} < T + T_H$): 
    
    The absorbed baby universe comes from a wormhole, with twist factor $s= 2\sqrt{E}(T_{\mathrm{eff}} - T) $.
    \item[(iii)] Absorbing branch ($T_{\mathrm{eff}} > T + T_H$):
    
    The absorbed baby universe comes from D-brane, with twist factor $s= 2\sqrt{E}T_H $.
\end{itemize}

\item For $T > T_H $, there are four cases (see Fig.\ref{T-S}(b)):

\begin{itemize}
    \item[(i)] Emitting branch ($0 < T_{\mathrm{eff}} < T - T_H$): 
    
    The emitted baby universe ends on D-brane, with twist factor $s=2\sqrt{E}T_H $.
    \item[(ii)] Emitting branch ($ T - T_H < T_{\mathrm{eff}} < T$): 
    
    The emitted baby universe comes from a wormhole, with twist factor $s= 2\sqrt{E}(T - T_{\mathrm{eff}}) $.
    \item[(iii)] Absorbing branch ($ T < T_{\mathrm{eff}} < T + T_H$): 
    
    The absorbed baby universe comes form a wormhole, with twist factor $ 2\sqrt{E}(T_{\mathrm{eff}} - T) $.
    \item[(iv)] Absorbing branch  ($T_{\mathrm{eff}} > T + T_H$): 
    
    The absorbed baby universe comes form D-brane, with twist factor $s= 2\sqrt{E}T_H $.
\end{itemize} 
\end{itemize}
All together , the probability of black hole is:

\begin{itemize}
    \item $T<T_{H}$:
    \begin{align}
        \mathcal{P}_{\mathrm{BH}} (T)=&\int_{0}^{\infty} dT_{\mathrm{eff}} \delta(T_{\mathrm{eff}}-T) +
        \nonumber\\
        &\frac{1}{T_H^2}\left[ \int_{0}^{T} dT_{\mathrm{eff}}   (T-T_{\mathrm{eff}}) + \int_{T}^{T+T_H} dT_{\mathrm{eff}}   (T_{\mathrm{eff}}-T)+\int_{T+T_H}^{\infty} dT_{\mathrm{eff}} T_H \right],
        \nonumber\\
        &=1-\frac{T}{T_H}+\frac{T^2}{2T_H^2}+\mathrm{const}.
        \label{BH-prob}
    \end{align}

    \item $T>T_{H}$:
    \begin{align}
         \mathcal{P}_{\mathrm{BH}} (T)=&\int_{0}^{\infty} dT_{\mathrm{eff}} \delta(T_{\mathrm{eff}}-T) +
         \nonumber\\
        & \frac{1}{T_H^2}\Bigg{[} \int_{0}^{T-T_H} dT_{\mathrm{eff}}  T_H + \int_{T-T_H}^{T} dT_{\mathrm{eff}}   (T-T_{\mathrm{eff}})+
        \nonumber\\
        &\int_{T}^{T+T_H} dT_{\mathrm{eff}}  (T_{\mathrm{eff}}-T)+\int_{T+T_H}^{\infty}  dT_{\mathrm{eff}} T_H  \Bigg{]}
        \nonumber\\
        &=\frac{1}{2}+\mathrm{const}.
    \end{align}
\end{itemize}
The constant term here is the same as for the white hole case, which is normalized to zero. 
    
\vspace{5pt}
\paragraph{Final results}

In summary, after regularization, our result for $\mathcal{P}_{\mathrm{WH}}(T)$ and $\mathcal{P}_{\mathrm{BH}}(T)$  is as follows:
\begin{align}
   &\textbf{For }\mathbf{ T< T_H}:\quad \mathcal{P}_{\mathrm{BH}} (T)= 1-\frac{T}{T_H}+\frac{T^2}{2T_H^2}, \quad \mathcal{P}_{\mathrm{WH}} (T)= \frac{T}{T_H}-\frac{T^2}{2T_H^2};
   \label{final-result1}
     \nonumber\\
 &\textbf{For }\mathbf{ T \ge T_H}:\quad \mathcal{P}_{\mathrm{BH}} (T)= \frac{1}{2}, \quad \mathcal{P}_{\mathrm{WH}} (T)= \frac{1}{2};
\end{align}


Our final result matches exactly with \cite{Blommaert:2024ftn} using the matrix integral. This agreement should not be surprising, since we encode the necessary non-perturbative  information into the coarse-graining procedure as we described above.

\vspace{5pt}
\subsection{Further comments}
Below we give several comments on our result:

We see that only the disk and handle-disk geometries with a twist factor cutoff are sufficient to recover the full non-perturbative result. Higher topologies are never important in our treatment\footnote{For a directly calculatio of higher genus, see \cite{Zolfi:2024ldx}.}, all the non-perturbative effects are encoded in the twist factor cutoff, which in turn is a result of treating the low lying degrees of freedom in gravity as FZZT branes.  
In \cite{Saad:2022kfe} the authors used the  genus expansion in the $\tau$-scaling limit  to recover the plateau. In their treatment higher genus contributions are crucial since they encode the non-perturbative information that leads to the plateau. This does not contradict our treatment which encodes the same non-perturbative effects through FZZT branes.

In our analysis, the Cauchy slice can emit or absorb a baby universe. These two processes comes from the same Euclidean geometry, the handle-disk, which should contribute in a time-symmetric way after  analytic continuation to Lorentzian time. As a result, the length of Cauchy slice can also be increased by absorbing a baby universe process. This situation differs somewhat from the analysis in \cite{Stanford:2022fdt}, where the authors proposed a "no short-cut" criterion claiming that, in order to correctly identify the interior, distant points along spatial slice should not be close together through the spacetime geometry. Their criterion implies that the correct Cauchy slice should always be the shortest one. However, our result is not in contradiction with theirs -- \cite{Stanford:2022fdt} focuses on the partition function of JT gravity, where careful analysis of  the mapping class group  is needed to avoid over-counting. The "no short-cut" criterion is designed to avoid such over-counting. In our case, however, we are calculating two-point function, where all geodesics in the bulk should be summed over. As noted in \cite{Saad:2019lba}, such summation precisely cancels out the constraints from the mapping class group, leaving the size of baby universe and twist factor unconstrained. Therefore, the "no short-cut" criterion does not apply to our calculation.

The non-perturbative twist factor cutoff is responsible for normalizing the sum of black hole and white hole probabilities to one at any time, and ensures the saturation to fifty-fifty probability after the Heisenberg time. Geometrically, the saturation comes from the fact that baby universes that lead the Cauchy slice to tunnel into white holes will be large and then necessarily end on D-branes.

In (\ref{final-result1}) we see that the handle-disk topology yields the quadratic term $\frac{T^2}{2T_H^2}$ in the final result in (\ref{final-result1}), while the twist factor cutoff gives the linear term $\frac{T}{T_H}$ contribution. This explains why the handle-disk is sufficient to reproduce the $\frac{1}{2}$ probability at the Heisenberg time, as was done by Stanford and Yang \cite{Stanford:2022fdt}. 

It is noteworthy that, in the final expression of black hole probability (\ref{final-result1}), the negative term $-\frac{T}{T_H}$ only comes from contributions where the baby universes end on the D-brane in the absorbing branch. However, this does not imply that D-brane somehow decreases the probability for the Cauchy slice to stay in black hole region. Instead, this term appears simply because as $T$ gets larger, the integration region  [$T+T_H$, $\infty$] narrows. Similarly, the negative term $-\frac{T^2}{2T_H^2}$ in the white hole probability arises also because the integration region $[T-T_H,0]$ shrinks as $T$ grows (before the Heisenberg time.)

\begin{figure}[htp!]
\begin{tabular}{cc}
    \begin{minipage}{0.5\hsize}
     \begin{center}
     \includegraphics[width=5cm]{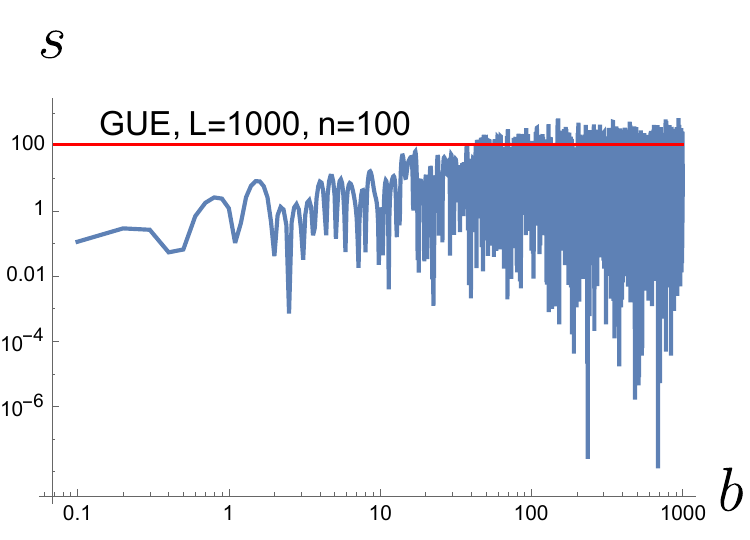}
    \end{center}
\end{minipage}
 \begin{minipage}{0.5\hsize}
     \begin{center}
     \includegraphics[width=5cm]{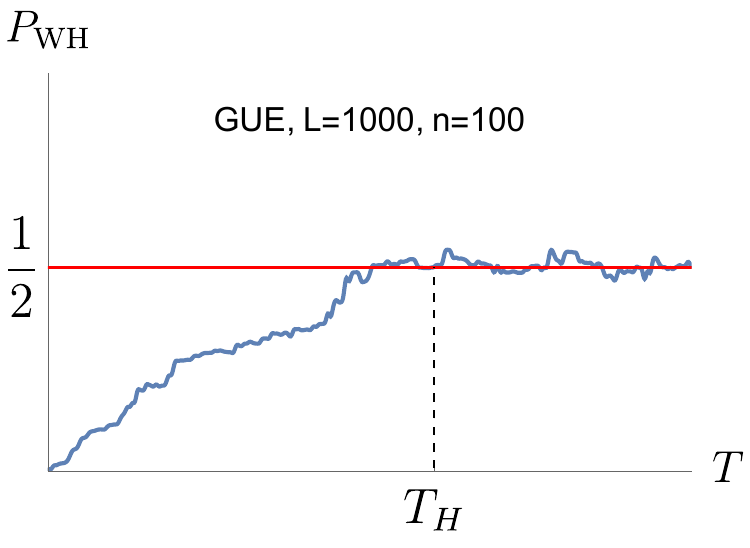}
    \end{center}
\end{minipage}
\\
\begin{minipage}{0.5\hsize}
     \begin{center}
     \includegraphics[width=5cm]{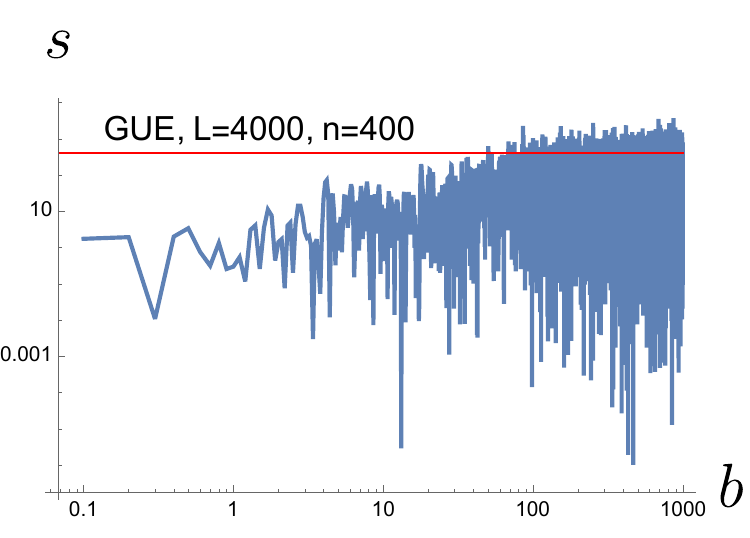}
    \end{center}
\end{minipage}
\begin{minipage}{0.5\hsize}
     \begin{center}
     \includegraphics[width=5cm]{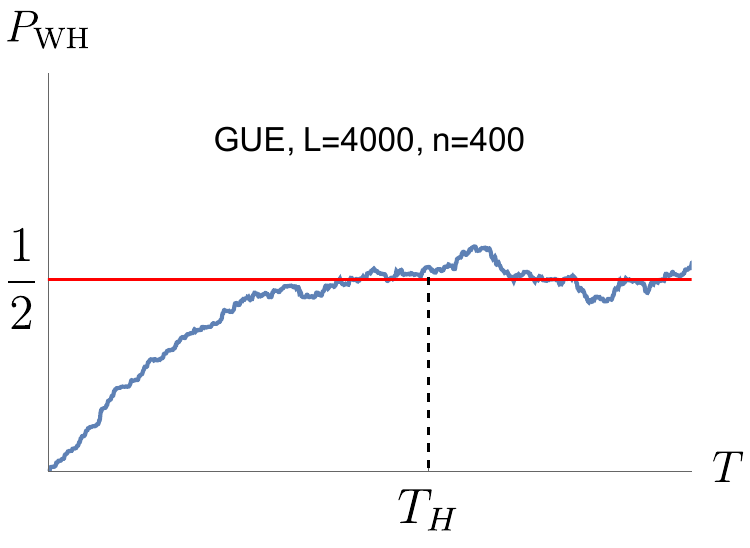}
    \end{center}
\end{minipage}
\\
\begin{minipage}{0.5\hsize}
     \begin{center}
     \includegraphics[width=5cm]{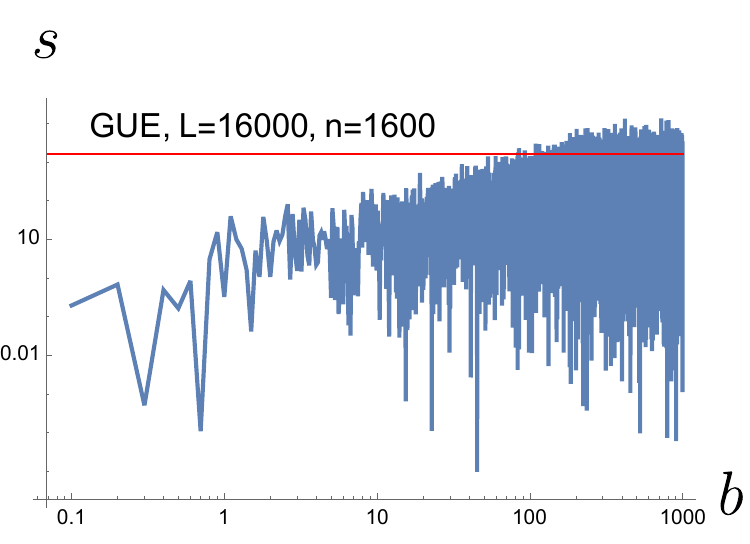}
    \end{center}
\end{minipage}
\begin{minipage}{0.5\hsize}
     \begin{center}
     \includegraphics[width=5cm]{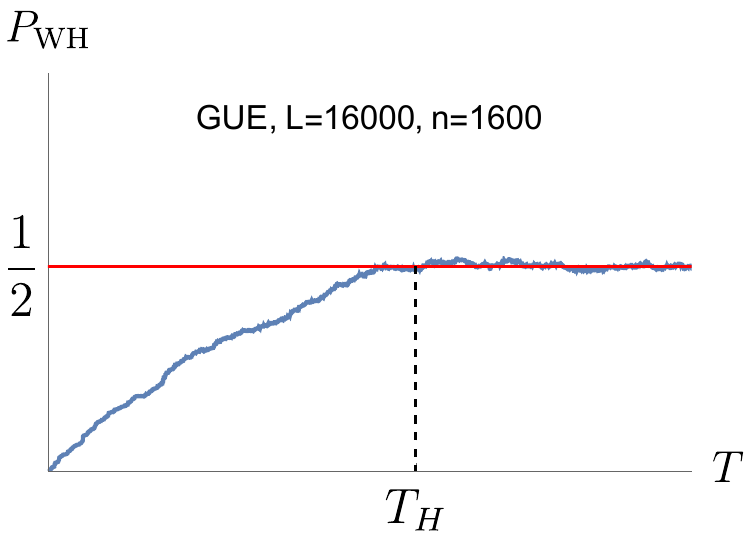}
    \end{center}
\end{minipage}
\\
\begin{minipage}{0.5\hsize}
     \begin{center}
     \includegraphics[width=5cm]{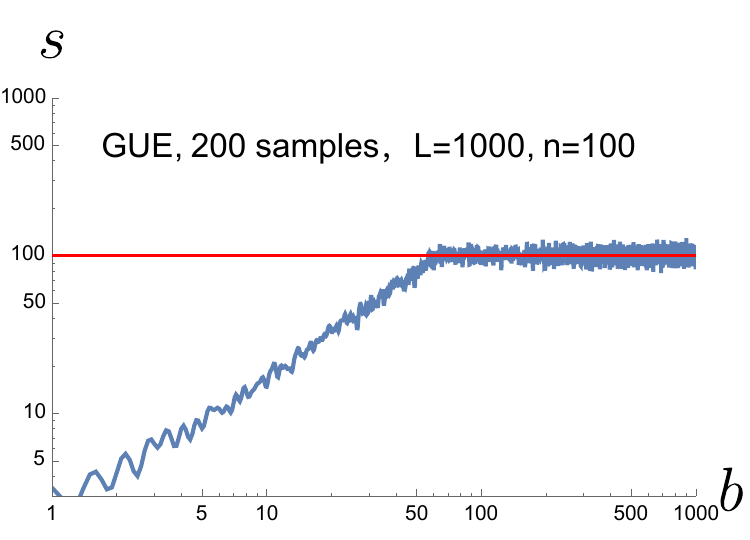}
    \end{center}
\end{minipage}
\begin{minipage}{0.5\hsize}
     \begin{center}
     \includegraphics[width=5cm]{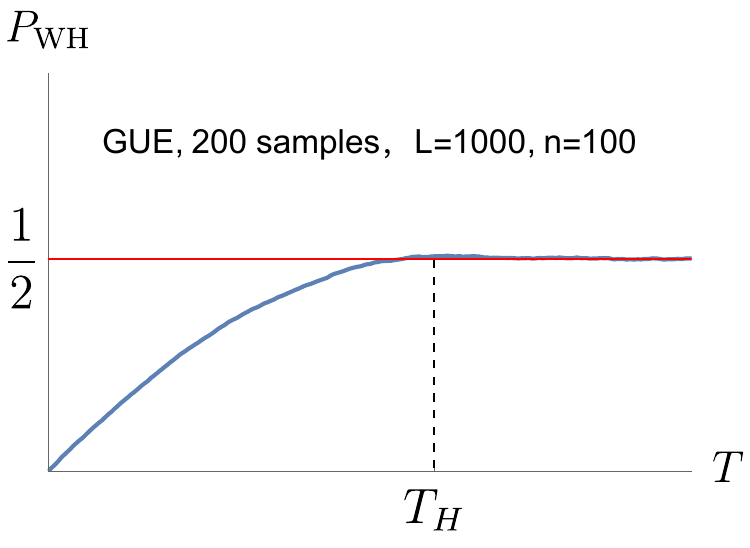}
    \end{center}
\end{minipage}
\end{tabular}
\caption{Left column: Numerical results for the twist factor s as a function of the baby universe size b.
Right column: Numerical results for the white hole probability $\mathcal{P}_{\mathrm{WH}}$ with respect to time $T$.
Here, $L$ is the total dimension of GUE matrices, and $n$ denotes the number of fixed (consecutive) eigenvalues drawn from the L$\times$L GUM matrix far from the edge. The first six diagrams display results form single samples, while the last two diagrams showed averaged results over 200 samples. We see that as $L$ and $n$ increase (right column of first six figures), the white hole probability approaches the averaged result, even though the corresponding input --- the twist factor -- still exhibits significant oscillations (left column of first six figures). }
\label{self-ave-prob}
\end{figure}

 \paragraph{Self-averaging property of the firewall probability.} Recall that our effective description involves integrating over configurations of $n$ FZZT branes, after forming an effective theory for these branes by averaging over the fast modes. Suppose we fix a single configuration of these branes, then we may expect the probabilities to be in a black or white hole states to exhibit heavy oscillations, similar to the oscillations of the SFF for a typical draw. This is not the case -- 
this probability involves an integration over the twist factor which acts to smooth the oscillations. As a result we find that the probabilities discussed here are self-averaging in this sense. In the large  $n$  limit, the probability approximates the coarse-grained result (\ref{final-result1}). We note that the self-averaging property holds if we fix all eigenvalues and go back to the UV description of a single member of the ensemble. This self-averaging behavior is demonstrated numerically in Fig.\ref{self-ave-prob}.

\vspace{10pt}
\section{Including matter loops}
\label{matter-loop_sec}

In this section we examine another potential source of firewalls by adding matter loops to our previous discussion, involving the  handle-disk geometry with the twist factor cutoff incorporating the effects of FZZT branes. This question was raised in \cite{Blommaert:2024ftn} -- particle-antiparticle pairs created through matter loops on the wormhole geometry may produce dangerous shock waves. Here we will argue qualitatively that such firewall effects are suppressed by at least $\mathcal{O}(1/T)$ at late times when compared to the firewall caused by wormhole shortening mechanism and thus only provides sub-dominant contribution.

We start by discussing the handle-disk without the twist factor cutoff, i.e. when the baby universe exchanged is sufficiently small. The one-loop contribution of a scalar with scaling dimension $\Delta$ is given by \cite{Jafferis:2022wez}:
\begin{align}
    Z_{\Delta}^{\mathrm{1-loop}}(b)&=\prod\limits_{n=0}^{\infty} \frac{1}{1-e^{-b(\Delta+n)}}, 
    \nonumber \\
    &=\sum\limits_{n=0}^{\infty}\frac{e^{-nb\Delta }}{(1-e^{-b})(1-e^{-2b})...(1-e^{-nb})},
    \nonumber\\
    &=\mathrm{exp}\left(\sum\limits_{n=1}^{\infty}\frac{e^{-nb\Delta }}{n(1-e^{-nb})}\right),
    \label{exp_div}
\end{align}
where $b$ is the size of wormhole throat. The first line is  obtained directly from usual CFT quantization on a cylinder, while the second line makes the contributions from primary and descendants states manifest, and the third line shows the divergence at small b.

\begin{figure}[htbp]
  \begin{center}
   \includegraphics[width=6cm]{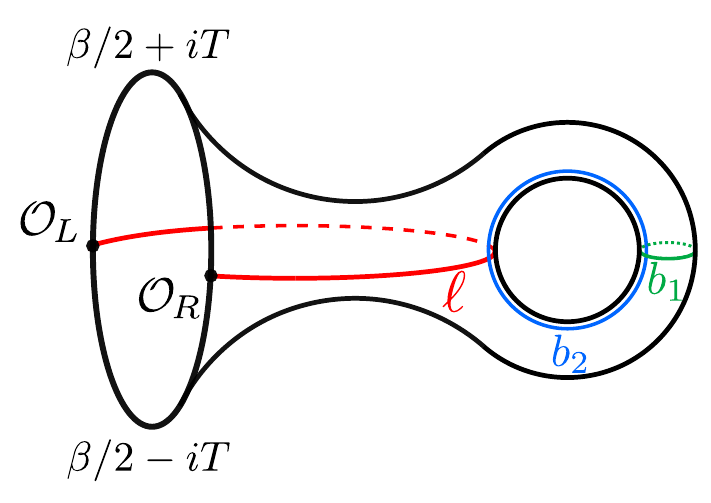}
  \end{center}   
\vspace*{-0.5cm}
\label{2pt_loop}
\end{figure}

There are two cycles where one can add a matter loop, corresponding $b_1$ and $b_2$ depicted in Fig.\ref{2pt_loop}. We now discuss them separately.

\vspace{5pt}
\subsection{Matter loop on \texorpdfstring{$b_1$ circle}{\texttt{psi}}}

Adding the matter loop at the cycle $b_1$ has no bearing on the firewall probability, since the entire loop remains inside the baby universe and has no contact with the Cauchy slice we denote by $\ell$ and therefore it can not produce shock waves or firewalls. Rather, it simply contributes as a one-loop determinant for the two-point function, which exhibits the well-known divergence at small $b_1$, as can be seen from (\ref{exp_div}). This divergence should be resolved through proper renormalization of fluctuations of the non-trivial topology but this divergence and its renormalization has no effect on the physics governing the emission or absorption of  baby universes and the firewall probability (for example it does not depend on the Loretzian time).


We can estimate the finite part of this one-loop determinant by setting $b_1$ to its value in the dominant geometry that contributes to the two-point function. Referring to (\ref{2pt_final}), the dominant contribution comes from $\ell_{T_\mathrm{eff}}\approx0$. Combining with $\ell_{T_\mathrm{eff}}=2|\sqrt{E}T_{\mathrm{eff}}|-\ln 4E$, the value of $b_1$ is:
\begin{align}
    b_1=2\sqrt{E}(T-T_{\mathrm{eff}})=2\sqrt{E}T-\ln4E \approx 2\sqrt{E}T \quad \text{for large T}.
\end{align}
Thus the correction of matter loop at $b_1$ at late time can be approximated as:
\begin{align}
    G_{\Delta}^{\mathrm{1-loop}}(T)\approx\mathrm{exp}\left(\sum\limits_{n=1}^{\infty}\frac{e^{-2\sqrt{E}Tn\Delta }}{n(1-e^{-2\sqrt{E}Tn})}\right)\times G_{\Delta}(T),
\end{align}
where $G_{\Delta}(T)$ is the previous two-point function in (\ref{2pt_final}) without the matter loop.

\vspace{5pt}
\subsection{Matter loop on \texorpdfstring{$b_2$}{\texttt{psi}} circle}

\begin{figure}[htbp]
    \centering
    \subfigure[]{\includegraphics[width=0.4\textwidth]{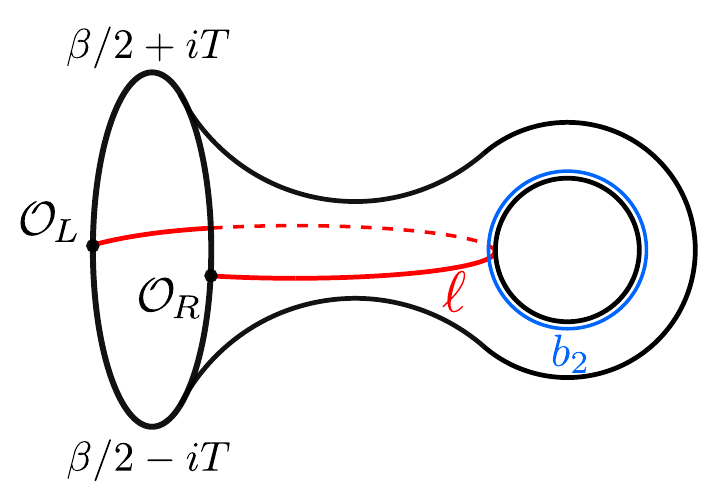}} 
    \subfigure[]{\includegraphics[width=0.5\textwidth]{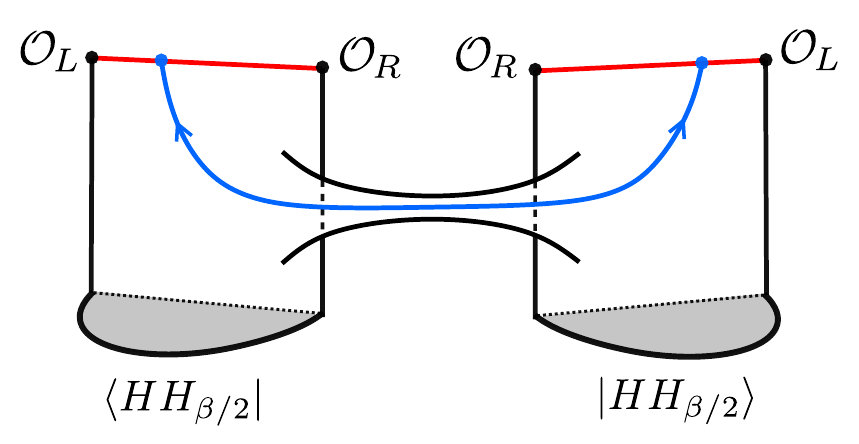}} 
    \caption{(a) Euclidean picture. (b) Lorentzian picture. }
    \label{firewall_loop}
\end{figure}

Adding matter loop at the cycle $b_2$ is more subtle and relevant, since it intersects with the Cauchy slice $\ell$ and thus can potentially create a dangerous shock wave. Indeed, in the Lorentzian picture depicted in fig.\ref{firewall_loop}, where we analytically continue the Euclidean geometry along the time symmetric Cauchy slice  $\ell$, a wormhole sends a particle through the matter loop $b_2$. If the particle is emitted at early enough time it experiences a large boost and may lead to a dangerous shockwave disrupting the Cauchy slice at late time. 

However, unlike the vacuum handle-disk geometry,  in this case the twist factor for the $b_1$ circle is no longer a zero mode due to the presence of the matter loop on the $b_2$ circle.\footnote{We are grateful to Zhenbin Yang for pointing out an error in our original analysis in this subsection. Much of the following analysis is based on discussions with him.} Indeed, twisting around $b_1$ further increases the length of the matter loop and will therefore be exponentially suppressed. As a result, at late times where the twist factor gives linear in $T$ contribution without the matter loop, the matter loop contribution now becomes subdominant, suppressed by $\mathcal{O}(1/T)$.



\begin{figure}[htbp]
  \begin{center}
   \includegraphics[width=6cm]{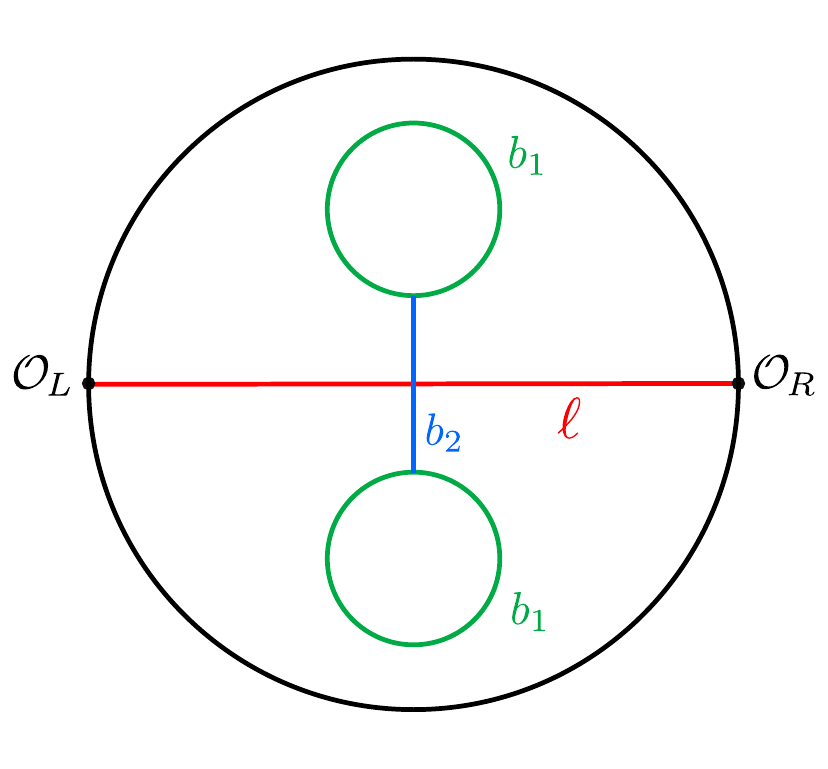}
  \end{center}   
\vspace*{-0.5cm}
\caption{}
\label{OTOC1}
\end{figure}

We begin by demonstrating the effective OTOC contour induced by the matter loop on the cycle $b_2$, which confirms the possibility for a firewall in the bulk.
The handle-disk geometry can be effectively described as a three-hole sphere by cutting $b_1$ open. This three-hole sphere can then be presented on the hyperbolic disk as in Fig.\ref{OTOC1}.

\begin{figure}[htbp]
  \begin{center}
   \includegraphics[width=6cm]{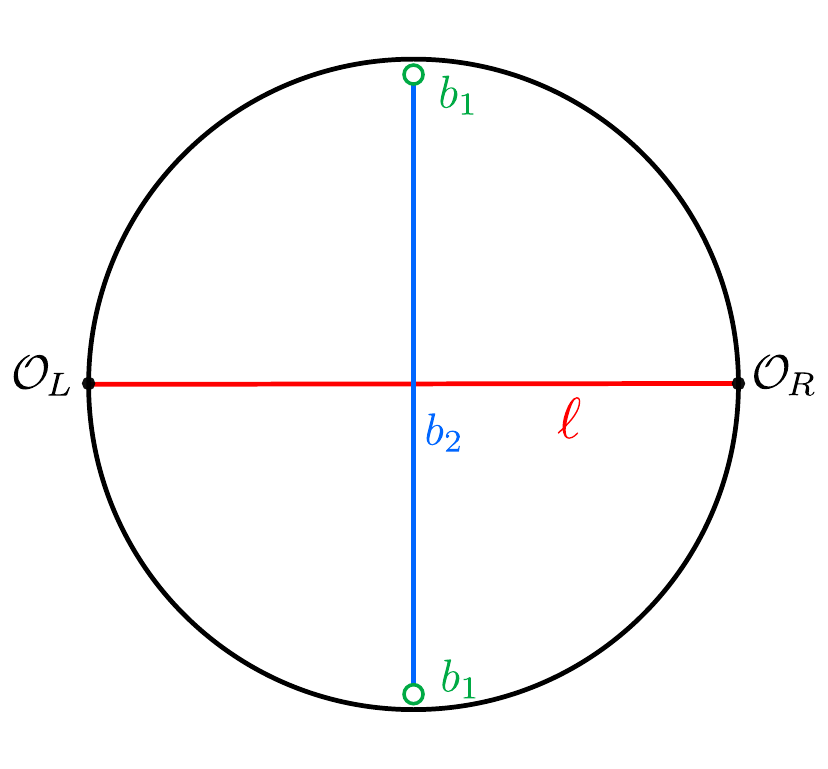}
  \end{center}   
\vspace*{-0.5cm}
\caption{}
\label{OTOC2}
\end{figure}

As the baby universe on the $b_1$ cycle approaches the boundary, the matter loop on $b_2$ effectively behaves like a pair of operator insertions at the boundary. Combined with the external operators $\mathcal{O}_L$ and $\mathcal{O}_R$, these insertions form an effective OTOC contour. This situation is depicted in Fig.\ref{OTOC2}\footnote{On this OTOC contour, the effective boundary time for the matter loop is governed by another twist mode which rotates the $b_2$ geodesic around the center of hyperbolic disk. This twist mode is not manifest in the current representation of handle-disk geometry using $b_1$ $b_2$ and $\ell$. A more useful representation for capturing this twist mode is demonstrated in Fig.4 of  \cite{Stanford:2021bhl}, where this twist mode corresponds to the $\theta$ coordinate after SL(2,R) gauge fixing. }.

However,  the above OTOC contour is sub-dominant at late times. It will be suppressed by the following factors:
\begin{itemize}
    \item In Fig.\ref{OTOC2}, since the matter loop can be effectively described as a pair of boundary operator insertions, part of its contribution involves the disk two-point function $e^{-\Delta\tilde{b}_2}$ where $\tilde{b}_2$ is the renormalized length as in (\ref{disk-2pt}). Since the twist mode $s_1$ around $b_1$ increases $b_2$ to $b_2+s_1$, this twist mode is now suppressed by a factor $e^{-\Delta s_1}$ and will be localized around zero. 
    
In contrast, in the  vacuum handle-disk case (\ref{2pt_final}),  this twist mode was a zero mode that is equal to the size of baby universe $b_1$. At late time (before plateau), this twist mode contributes a linear in $T$ term.
    
Thus, compared to the vacuum handle-disk, the matter loop case is  suppressed by $\mathcal{O}(1/T)$ at late times due to the localization of the twist mode.
    
    \item On the dominant geometry at late times, the moduli parameters are set as follows:
        \begin{align}
    b_1=2T\sqrt{E}-\ln 4E,  \quad  b_{2}=\mathrm{arccosh}(9), \quad \ell=0.
    \label{hd_saddle}
    \end{align}
   The detailed derivation is given in Appendix.\ref{appen_loop det}. For these values of the moduli, $b_2$ is $\mathcal{O}(1)$ and the Cauchy slice $\ell$ only probes a thermal $b_2$ loop with $\mathcal{O}(1)$ temperature. 

  
To generate a strong shock wave in the bulk, $b_1$ must approach to the OTOC contour shown in Fig.\ref{OTOC2}. This indicates that the firewall moduli are expected to deviate a lot from the semi-classical saddle (\ref{hd_saddle}). Compared with  vacuum handle-disk case, where the saddle (\ref{hd_saddle}) lies at the threshold for creating a firewall-such that any  $b_1> 2\sqrt{E}T$ could produce one, the probability of a firewall in this case is expected to be further suppressed due to the reduced moduli region.\footnote{When $b_1$ approaches the boundary, the semiclassical value of $b_2$ will be determined by the relative boost between the effective boundary time of matter loop and $\ell$ (after trading of baby universe). A small boost might results in a small value of $b_2$ due to Lorentzian cancellation. This complicates the identification of moduli space.  It would be interesting to make this point more precise,  which we leave for future work. } 
  
\end{itemize}

As a result, the firewall caused by the matter loop can be ignored at late times. The most significant case of a firewall still arises from the wormhole shortening mechanism discussed in Sec.\ref{grayhole_sec}.

\vspace{5pt}
\subsection{Large baby universes}

The discussion thus far is relevant for matter loops in the presence of sufficiently small baby universes, such that the effective gravitational description is a wormhole. For completeness we can discuss the matter loop for large baby universes which end on one of our FZZT branes. 

We argue that aside for the treatment of the twist factor the conclusions are identical. To this end we use the picture of correlated effective branes, resulting from the integration over the FZZT branes, which are found to be  maximally entangled (see (\ref{entangle_BU})). We can therefore define the matter loop as the probe matter field operator acting on these maximally entangled states,
\begin{align}
    |\delta\Delta\rangle\equiv 
    \int_0^{\infty} s(b) ~db~\delta\mathcal{O}_{\Delta}\cdot 
    \Big{(}|b\rangle\otimes|b\rangle \Big{)}.
\end{align}
The maximal entanglement in the large $b$ basis ensures that the matter loop carries through smoothly, resulting effectively in a matter loop, whether we are in the smooth wormhole regime or the correlated branes, both contain a delta function (see (\ref{entangle_BU})) restricting to $b_1=b_2$. The only difference is the treatment of the twist factor $s(b)$. 
Thus our previous statement still applies, the only modification is that the effect of matter loops after Heisenberg time will be suppressed at least by $\mathcal{O}(1/T_H)$.

\vspace{10pt}
\section{Summary and future directions}

We use the framework of JT gravity together with FZZT branes as as an effective Wilsonian description of the matrix integral, as described in in Sec.\ref{FZZT_sec}. This provides a natural framework of encoding non-perturbative effects long time effects into the genus expansion.  This approach allows us to complete the wormhole shortening picture with the full non-perturbative corrections summarized  by an effective twist factor with a cutoff. The effective twist factor separates the D-brane and wormhole contributions at any time and preserves a clear geometric picture. 

By applying this twist factor cutoff prescription to the two-point function on the handle-disk geometry, we find that the white hole probability increases at late time and eventually saturates to $1/2$ after the Heisenberg time. This validates the gray hole conjecture and indicates a 1/2 possibility for an in-falling observer to encounter a firewall behind horizon at late time. This result remains valid even for a single quantum system without ensemble averaging. Meanwhile,
through the twist mode analysis, another potential source of firewall induced by the matter loop is at least $\mathcal{O}(1/T)$ suppressed at late times.

There are several interesting future directions. The notion of firewall should be quantified more precisely bu constructing a precise model for an infalling observer, perhaps along the lines of \cite{Jafferis:2022wez,Gao:2021tzr,deBoer:2022zps}. The twist factor cutoff can be also used for other dilaton gravity model that do not have a clear matrix dual. In particular it would be interesting to examine the finite $\sigma$ setup discussed in \cite{Blommaert:2021gha}. This corresponds to the situation where only a few FZZT branes are inserted into the micro-canonical windows.   In this case, we might expect that the double trumpet and two-point functions to undergo a "tearing phase" transitions with extra macroscopic holes \cite{Kazakov:1989cq}.  
In such case, the two-point function lacks a clear genus expansion, making it difficult to maintain a clear geometric picture. However, it may still be possible to encode the dual deformed matrix integral results into the effective twist factor. On the gravity side, we can still use the handle-disk geometry with this effective twist factor to calculate the firewall probability. It would be interesting to check if how  the firewall probability might be modified during this tearing phase transition. 


    
    

\vspace{10pt}
\acknowledgments
We would like to specially thank Zhenbin Yang for valuable discussions and pointing out an error in an early version of this paper. We also thank  Sean McBride, Jeremy van der Heijden and Abhisek Sahu for helpful discussions. Our work is funded by a Discovery grant from NSERC.

\vspace{10pt}
\begin{appendix}

    \section{Semiclassical approximation of \texorpdfstring{$\psi_{E+\frac{\omega}{2}}(\ell) \, \psi_{E-\frac{\omega}{2}}(\ell)$}{\texttt{psi}}}
    \label{appen_semi}

    In this appendix we derive the approximation of the expression  $ \psi_{E+\frac{\omega}{2}}(\ell) \psi_{E-\frac{\omega}{2}}(\ell)$ in micro-canonical ensemble with large $E$, following \cite{Blommaert:2024ftn}. First we express the wavefunction in terms of the modified Bessel functions and use the integral representation:
\begin{equation}
    K_{i\alpha}(x)=\frac{1}{2}\int_{-\infty}^{\infty} db e^{-x\mathrm{cosh}b}e^{-i\alpha b}.
\end{equation}
Giving,
\begin{align}
    \psi_{E+\frac{\omega}{2}}(\ell) \psi_{E-\frac{\omega}{2}}(\ell) &= 2 K_{2i \sqrt{E-\frac{\omega}{2}}}\left(2e^{-\ell/2} \right) 2 K_{2i \sqrt{E+\frac{\omega}{2}}}\left(2e^{ -\ell/2} \right),  \nonumber\\
    &=\int_{-\infty}^{\infty} db_1 db_2 e^{-2e^{-\frac{l\ell}{2}} \mathrm{cosh}(b_1)} e^{-2e^{-\frac{\ell}{2}} \mathrm{cosh}(b_2)} e^{-2i(E-\frac{\omega}{2})^{\frac{1}{2}}b_1} e^{-2i(E+\frac{\omega}{2})^{\frac{1}{2}}b_2}, \nonumber\\  
    &\approx \int_{-\infty}^{\infty} db_1 db_2 e^{- 2 \left( e^{-\frac{\ell}{2}}\mathrm{cosh}(b_1)+e^{-\frac{\ell}{2}}\mathrm{cosh}(b_2)\right)} e^{-2iE^{1/2}(b_1+b_2)} e^{-2i\frac{\omega}{4E^{1/2}}(b_2-b_1)}.
    \label{semi-wave}
\end{align}

Relabeling the variables,
\begin{align}
    \pi-2ib_1=2iE^{1/2} (G+g), \quad  \pi-2ib_2=2iE^{1/2} (-G+g).
\end{align}
the integral (\ref{semi-wave}) becomes:
\begin{align}
    2E\int_{-\infty}^{\infty} dG dg \exp{\bigg \{-2\pi E^{1/2}+4iEg - i \omega G-i2\left[e^{-\frac{\ell}{2}}\mathrm{sinh}\left(E^{\frac{1}{2}}(G+g)\right)+e^{-\frac{\ell}{2}}\mathrm{sinh}\left(E^{\frac{1}{2}}(-G+g)\right)\right]\bigg\}}.
\end{align}

The saddle point equations of $G$ and $g$ for this integral are:
\begin{align}
    &\frac{\partial S}{\partial g}\Bigg |_{g=g_{\star}} =0, \quad \frac{\partial S}{\partial G}\Bigg |_{G=G_{\star}} =0 \quad \Rightarrow \nonumber \\
    &(4E-\omega)=4E^{1/2}e^{-\frac{\ell}{2}}\mathrm{cosh}(E^{1/2}(G_{\star}+g_{\star})), \quad  (4E+\omega)=4E^{1/2}e^{-\frac{\ell}{2}}\mathrm{cosh}(E^{1/2}(-G_{\star}+g_{\star})),
\end{align}
with two solutions ($\omega \rightarrow 0$):
\begin{align}
    G_{\star} &= \frac{ \mathrm{arccosh}(E^{1/2}e^{\frac{\ell}{2}})}{ E^{1/2}}, \quad 
    g_{\star} = 0; \nonumber \\
    &\quad \mathrm{or}: G_{\star}\rightarrow -G_{\star}, \quad g_{\star} \rightarrow -g_{\star}.
\end{align}
Note that the above solution is only valid when $E^{1/2}e^{\frac{\ell}{2}}\geq 1$. So we need to add Heaviside function  $\Theta \left(\log(E)+\ell \right)$ to the result.

 The on-shell action is obtained as:
\begin{align}
    S_{\star}=&-2\pi E^{1/2} \pm \frac{i\omega }{E^{1/2}} \mathrm{arccosh}(E^{1/2}e^{\frac{\ell}{2}}).
\end{align}

 One-loop determinant for this saddle point is given as:
\begin{align}
    \mathrm{det}\left( \begin{array}{cc} \frac{\partial^2 S}{\partial G \partial G} &  \frac{\partial^2 S}{\partial G \partial g} \\
\frac{\partial^2 S}{\partial g \partial G} & \frac{\partial^2 S}{\partial g \partial g} \end{array} \right) \Bigg |_{G=G_{\star}, g=g_{\star}}=16 E^2  ( E- e^{-\ell}).
\end{align}

All in all, including two sets of saddle points and one-loop determinant contributions, the wavefunction product can be approximated as:
\begin{align}
     \psi_{E+\frac{\omega}{2}}(\ell) \psi_{E-\frac{\omega}{2}}(\ell)
     \approx & \frac{ 2\pi }{\sqrt{ E- e^{-\ell}}} \exp{(-2\pi E^{1/2})} \mathrm{cos}
     \left(
     \frac{\omega}{E^{1/2}} \mathrm{arccosh}(e^{\frac{\ell}{2}}E^{1/2}) \right)
     \Theta \left(\log(E)+\ell \right), \nonumber\\
     \approx & \frac{ e^{S_0}}{2\pi \sqrt{ E- e^{-\ell}} \rho_{0}(E)} \mathrm{cos}
    \left( \frac{k}{E^{1/2}} \mathrm{arccosh}(e^{\frac{\ell}{2}}E^{1/2}) \right)   \Theta \left(\log(E)+\ell \right), 
     \label{semi-appro-precise}
\end{align}
where in the second approximation we used the fact that $E \gg 1$ so $\mathrm{exp}(-2\pi E^{\frac{1}{2}})\approx \frac{e^{S_0}}{4\pi^2 \rho_{0}(E)}$.

\vspace{10pt}
\section{One-loop determinant of the \texorpdfstring{$b_2$}{\texttt{psi}} circle}
\label{appen_loop det}

\begin{figure}[htbp]
  \begin{center}
   \includegraphics[width=8cm]{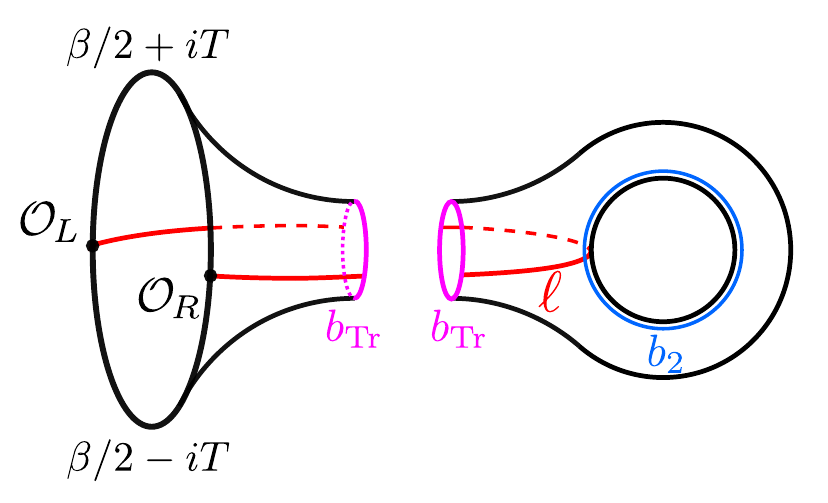}
  \end{center}   
\vspace*{-0.5cm}
\caption{}
\label{2pt_decomp}

\end{figure}

To calculate the one-loop determinant of $b_2$, it is more convenient to decompose the handle-disk geometry as shown in Fig.\ref{2pt_decomp}, into a trumpet with one asymptotic boundary and one geodesic boundary of size $b_{\mathrm{Tr}}$ and a pair of pants with three geodesic boundaries: $b_{\mathrm{Tr}}$, $b_2$, $b_2$, where we sum over all the values of $b_2$.

\begin{figure}[htbp]
  \begin{center}
   \includegraphics[width=7cm]{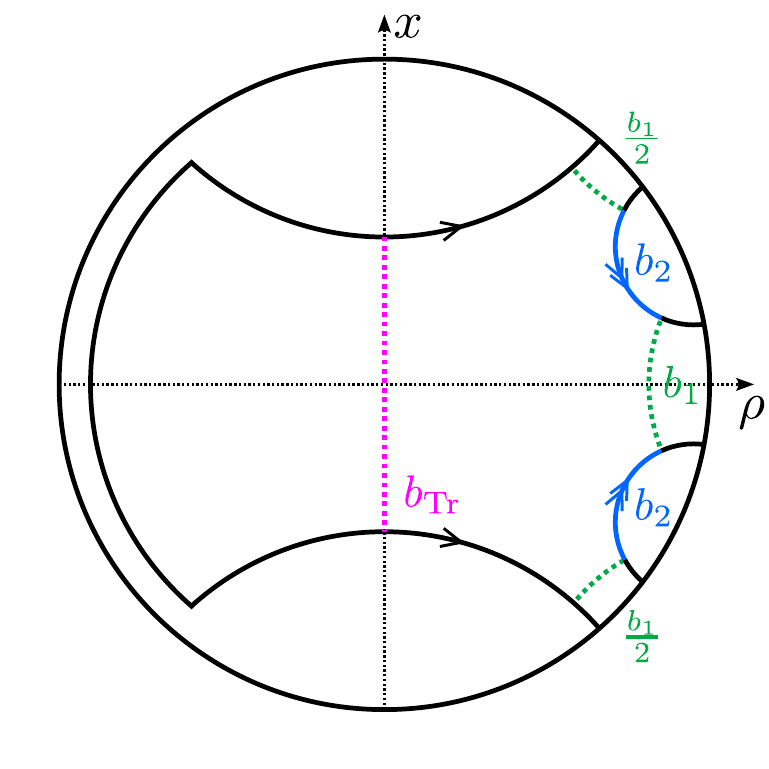}
  \end{center}   
\vspace*{-0.5cm}
\caption{}
\label{fig_quot}
\end{figure}

To perform the calculation, we present the handle-disk geometry as a quotient of the hyperbolic disk in Fig.\ref{fig_quot}. We work with the  Fermi coordinates:
\begin{equation}
    ds^2 = d\rho^2 + \mathrm{cosh}^2\rho dx^2,
\end{equation}
where the trumpet region has $\rho < 0$ (left hand side of $b_{\mathrm{Tr}}$)   and the pair of pants region has $\rho >0$  (right hand side of $b_{\mathrm{Tr}}$)  with two boundaries $b_1$ being glued together.

To find the value of $b_2$ after analytically continuing the two-point function to Lorentztian time $T$, we use the same method as in the $b_1$ case by matching Euclidean data to the dominant contribution at late times.  The case of $b_2$ is more complicated since it depends on both $b_1$ and $b_{\mathrm{Tr}}$. Let us now calculate the circle $b_2$ and the length of Cauchy slice $\ell$ in terms $b_1$ and $b_{\mathrm{Tr}}$ on this hyperbolic disk.

\vspace{5pt}
\subsection{Length of \texorpdfstring{$\ell$}{\texttt{psi}}}

The two boundary operators $\mathcal{O}_L$ and $\mathcal{O}_R$ are inserted at asymptotic boundary symmetrically with respect to the center of the coordinate system. Semi-classically, we can parametrize the coordinates of the boundary operators as:
\begin{align}
    \mathcal{O}_L : (-\rho_1, b_{Tr}/4), \quad \mathcal{O}_R : (-\rho_2, -b_{Tr}/4).
\end{align}
with $\rho_1, \rho_2 \gg1$.

The length of $\ell$ can be decomposed as follows:
\begin{align}
    \ell=\rho_1+\rho_2+d_1+d_2.
\end{align}
\begin{figure}[htbp]
  \begin{center}
   \includegraphics[width=7cm]{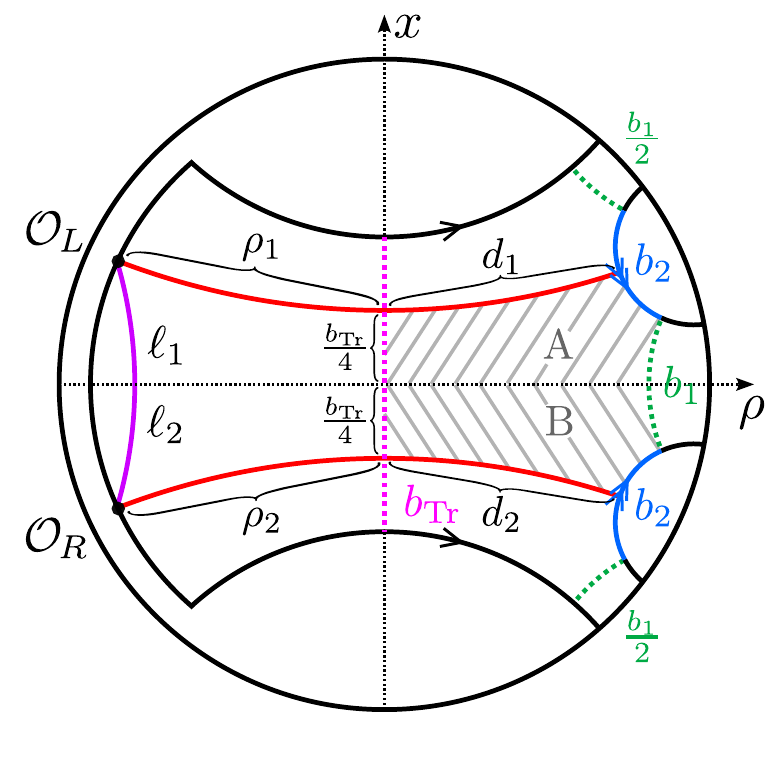}
  \end{center}   
\vspace*{-0.5cm}
\caption{}
\label{hyp_l}
\end{figure}
We can calculate $d_1$ and $d_2$ using trigonometry of the pentagons A and B in Fig.\ref{hyp_l} \cite{10.1007/978-0-8176-4992-0}:
\begin{align}
    \cosh\frac{b_1}{2}=\sinh\frac{b_{\mathrm{Tr}}}{4}\sinh d_1=\sinh\frac{b_{\mathrm{Tr}}}{4}\sinh d_2.
\end{align}
This gives the length of $d_1$ and $d_2$:
\begin{align}
    e^{d_1}=e^{d_2}= \frac{\cosh\frac{b_1}{2}}{\sinh \frac{b_{\mathrm{Tr}}}{4}}+\sqrt{\frac{\cosh^2\frac{b_1}{2}}{\sinh^2 \frac{b_{\mathrm{Tr}}}{4}}+1} 
\end{align}
Thus, the total length of $\ell$ is:
\begin{align}
    e^{\ell}=\left(\frac{\cosh\frac{b_1}{2}}{\sinh \frac{b_{\mathrm{Tr}}}{4}}+\sqrt{\frac{\cosh^2\frac{b_1}{2}}{\sinh^2 \frac{b_{\mathrm{Tr}}}{4}}+1} \right)^2 e^{\rho_1+\rho_2}
\end{align}

Meanwhile, we can also calculate the length of $\ell_1$ and $\ell_2$ in Fig.\ref{hyp_l} by  using trigonometry of the polygon bounded by $\ell_1(\ell_2), \rho_1(\rho_2), \frac{b_{\mathrm{Tr}}}{4}$ and the $\rho$ axis\footnote{We thank Zhenbin Yang for pointing out this constraint, which we had overlooked in the previous draft.}:
\begin{align}
    \sinh(\ell_1)&=\sinh(\frac{b_{\mathrm{Tr}}}{4})\cosh(\rho_1), \\
     \sinh(\ell_2)&=\sinh(\frac{b_{\mathrm{Tr}}}{4})\cosh(\rho_2).
\end{align}

The quantity $\ell_1+\ell_2$ represents the geodesic attaching between the two boundary operators without going through the handle. Thus, $\ell_1+\ell_2$  simply corresponds to the geodesic length on the disk, where the presence of the handle only contributes to the renormalization of disk area, which we denote as $\ell_D$. In $\rho_1, \rho_2 \gg1$ limit, $\ell_D$ is obtained as: 
\begin{align}
    \ell_D=\rho_1+\rho_2+2\log\sinh(\frac{b_{\mathrm{Tr}}}{4}).
\end{align}

\vspace{5pt}
\subsection{Length of \texorpdfstring{$b_2$}{\texttt{psi}}}

We can calculate $b_2$ via the following Hexagon on the hyperbolic disk:
\begin{figure}[htbp]
  \begin{center}
   \includegraphics[width=5cm]{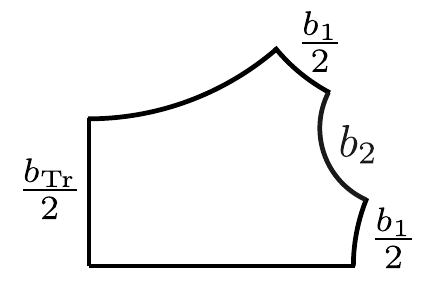}
  \end{center}   
\vspace*{-0.5cm}
\caption{}
\label{hex}
\end{figure}

The sides satisfy:
\begin{align}
    \cosh\frac{b_{\mathrm{Tr}}}{2}=\sinh^2\frac{b_1}{2}\cosh b_2 -\cosh^2 \frac{b_1}{2}.
\end{align}
So length of $b_2$ is obtained as:
\begin{align}
    \mathrm{cosh} b_2 =\frac{\mathrm{cosh}\frac{b_{\mathrm{Tr}}}{2}+1}{\mathrm{sinh}^2\frac{b_1}{2}}+1.
\end{align}

\vspace{5pt}
\subsection{One-loop determinant}

In sum, we have three related formulas:
\begin{align}
    &e^{\ell} = \left(\frac{\mathrm{cosh}\frac{b_1}{2}}{\mathrm{sinh}\frac{b_{Tr}}{4}}+ \sqrt{\frac{\mathrm{cosh}^2\frac{b_1}{2}}{\mathrm{sinh}^2\frac{b_{\mathrm{Tr}}}{4}}+1}  \right)^2e^{\rho_1+\rho_2}, \quad (\text{hyperbolic data})
    \label{f1} \\
      &\mathrm{cosh} b_2 =\frac{\mathrm{cosh}\frac{b_{\mathrm{Tr}}}{2}+1}{\mathrm{sinh}^2\frac{b_1}{2}}+1, \quad (\text{hyperbolic data}) 
       \label{f2}
      \\
      &\ell_D=\rho_1+\rho_2+2\log\sinh(\frac{b_{\mathrm{Tr}}}{4}),\quad (\text{hyperbolic data}) 
       \label{f3}
      \\
    &\ell =|2T\sqrt{E}\pm b_1|-\mathrm{ln}4E=0, \quad (\text{Lorentzian saddle of handle-disk})
     \label{f4}
    \\
    &\ell_D =2T\sqrt{E}-\mathrm{ln}4E. \quad (\text{Lorentzian saddle of disk})
     \label{f5}
\end{align}

Solving the above equations at large T limit, we find:
\begin{align}
    b_1=2T\sqrt{E}-\ln 4E,  \quad b_{\mathrm{Tr}}=2b_1+4\log2, \quad b_{2}=\mathrm{arccosh}(9).
\end{align}

 Thus, it turns out that the one-loop determinant on the circle $b_2$ is simply an $\mathcal{O}(1)$ factor depending on the scaling dimension $\Delta$ on the saddle geometry:
 \begin{align}
     Z_{\Delta}^{\text{1-loop}}(b_2=\mathrm{arccosh}(9))=\prod\limits_{n=0}^{\infty} \frac{1}{1-e^{-\mathrm{arccosh}(9)(\Delta+n)}} \approx\mathcal{O}(1).
 \end{align}

    \end{appendix}

\vspace{10pt}
\bibliographystyle{JHEP}
\bibliography{reference}

\end{document}